# Design Principles of Length Control of Cytoskeletal Structures

Lishibanya Mohapatra[1], Bruce L. Goode[2], Predrag Jelenkovic[3], Rob Phillips[4], Jane Kondev[5,*]

[1]Department of Physics, Brandeis University, Waltham, MA, USA

lishi87@brandeis.edu

[2]Department of Biology and Rosenstiel Basic Medical Sciences Research Center, Brandeis University, Waltham, MA, USA

goode@brandeis.edu

[3]Department of Electrical Engineering, Columbia University, New York, NY

predrag@ee.columbia.edu

[4]Department of Applied Physics and Division of Biology and Biological Engineering, California Institute of Technology, Pasadena, CA, USA

phillips@pboc.caltech.edu

[5]Department of Physics, Brandeis University, Waltham, MA, USA

kondev@brandeis.edu

* Corresponding author; Department of Physics, MS 057, Brandeis University, Waltham, MA 02454. Phone (781) 736-2812. Fax (781) 736-2915. E-mail kondev@brandeis.edu

*Shortened running title*

*Length Control of Cytoskeletal Structures*

**Table of contents**

*Abstract*: Cells contain elaborate and interconnected networks of protein polymers which make up the cytoskeleton. The cytoskeleton governs the internal positioning and movement of vesicles and organelles, and controls dynamic changes in cell polarity, shape and movement. Many of these processes require tight control of the size and shape of cytoskeletal structures, which is achieved despite rapid turnover of their molecular components. Here we review mechanisms by which cells control the size of filamentous cytoskeletal structures from the point of view of simple quantitative models that take into account stochastic dynamics of their assembly and disassembly. Significantly, these models make experimentally testable predictions that distinguish different mechanisms of length-control. While the primary focus of this review is on cytoskeletal structures, we believe that the broader principles and mechanisms discussed herein will apply to a range of other subcellular structures whose sizes are tightly controlled and are linked to their functions.

*1. Introduction*

A remarkable feature of all living cells is that they have a variety of distinguishable subcellular parts (organelles) with characteristic sizes and shapes. These structures have been observed since the dawn of microscopy and yet it is only recently that we have developed experimental tools to address key questions, such as: How do organelles obtain their specific shapes, and how do cells control their number and size? For example, how does a cell 'decide' how many mitochondria or centrioles should it have? Or, how does a cell construct structures with precisely arranged parts, such as sarcomeres in muscle with its regimented arrays of actin filaments interdigitated with myosin fibers? The cytoskeleton provides a particularly fruitful arena to develop quantitative models that address these questions of morphology, in light of the wealth of quantitative information about its structure and dynamics at the molecular level. In this review, we use theory as a guide and a common language for describing the various size control mechanisms that have been proposed recently for diverse cytoskeleton structures. By reviewing the field from the point of view of simple models we hope to identify fruitful directions for new experiments.

The cytoskeleton consists of a number of organelles and substructures that seem to be designed with a precise size and geometry, suggesting that these physical properties are intimately tied to their biological functions. Examples include cytoskeletal structures such as the mitotic spindle, actin cables, and cilia. The majority of cytoskeleton structures are comprised of protein polymers such as microtubules and actin filaments, which are themselves made up of simple building block proteins such as tubulin dimers and actin monomers, respectively. How these structures are able to maintain a remarkably constant size despite undergoing highly dynamic turnover of their components is still not well understood.

*1.1 Case studies of cytoskeletal structures*

In cells, we find numerous examples of cytoskeletal structures with sizes that are dictated by the particular cellular process they control. For example, during cell division, the microtubule-based mitotic spindle maintains a constant size during metaphase and is relatively constant in size within cells of a given type. For example, in *Drosophila* S2 cells, the length of the spindle at metaphase varies little from cell to cell (11.5 ± 2.0 microns)(18). Another example are cilia, which are microtubule-based structures used by all eukaryotic cells for motility and sensation (37, 38). In the cellular alga *Chlamydomonas* they grow to about 10 microns in length (2, 24, 34). In budding yeast cells, actin cables are used for intracellular transport and they grow to the approximate length of the mother cell ($\simeq$ 5 microns)(7, 52). Stereocilia are mechanosensory actin protrusions on the surface of sensory (hair) cells in the inner ear, and are the key players in the transduction of sound or motion into the electrical signals that underlie our senses of hearing and balance. Stereocillia can range from 10-120 microns, but in a given hair cell they are always graded in length to take on a characteristic staircase arrangement (32, 33).

While these cellular structures have lengths at the micron scale, the mechanisms controlling them operate at the nanometer scale, at the level of individual proteins interacting with each other. Many years of research in molecular and cell biology have revealed a detailed list of molecules involved in shaping the cytoskeleton, but how they all work together, i.e., what are the physical mechanisms that cells employ to achieve precise control of this subcellular structure, is still largely unknown.

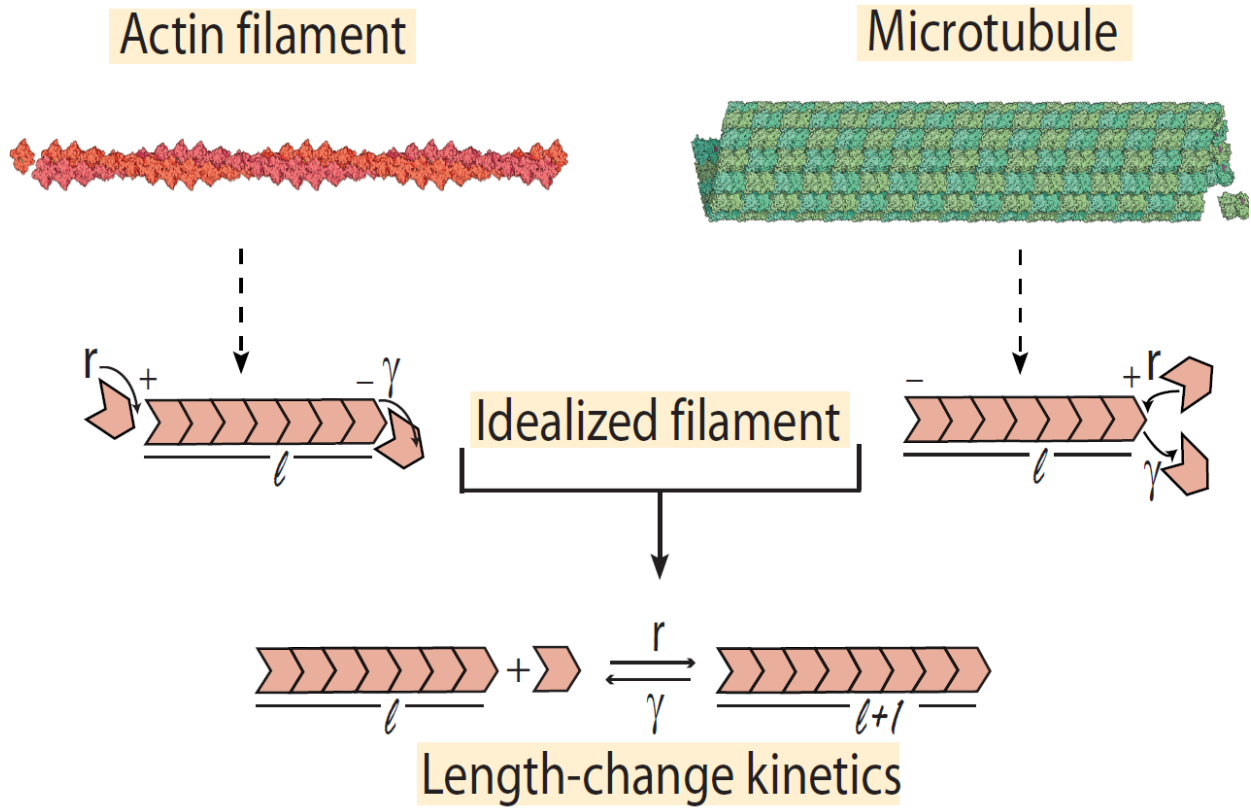

**Figure 1: Simple model of assembly dynamics of cytoskeletal filaments.** Microtubules and actin filaments are polar polymers comprised of individual monomers that can be added or removed from their ends. In actin filaments, monomers can be added to and subtracted from both the barbed (or plus) ends or pointed (or minus) ends. However, in vivo there is rapid addition of monomers at the barbed end, and net dissociation of monomers at the pointed end. In contrast, the addition and removal of monomers happens primarily at the plus end of microtubules. We abstract these two classes of cytoskeleton polymers into a simplified model by considering idealized filaments for which monomer subunits (orange) are added with rate $r$ and subtracted with rate $\gamma$. This results in the length ($l$) of the filament evolving stochastically in time.

*1.2 Length dependent assembly and disassembly rates*

In this review we study the various mechanisms that cells employ to control the size of filamentous structures made up of microtubules and actin filaments. Microtubules are hollow tubular polymers of tubulin. Their outer radius is about 24 nm and inner radius is about 12 nm. Actin filaments, on the other hand, are linear, helical two-stranded polymers of actin subunits, some 6 nm in diameter. In vivo, microtubules grow and shrink primarily from their plus ends, whereas actin filaments grow at their plus ends and shrink at their minus ends. Even though microtubules and actin filaments seem quite different, they also have important similarities; they all grow (or shrink) by adding (or removing) constituent subunits at their ends (**Figure 1**), and the nucleotide on the individual subunits are rapidly hydrolyzed when the subunit is added to the polymer end. Herein, we consider an abstraction in which each filamentous structure is viewed as a polymer filament made up of constituent monomers. We study the dynamics of the filament as monomers are added or subtracted from it, suppressing the rich and nuanced internal structure that characterizes real cytoskeletal filaments. In a sense, we do not make a distinction between actin-based and microtubules-based structures and thus our results are conceptually applicable to both.

If a particular length-control mechanism were to result in a structure of a specific length, we expect the corresponding growth trajectory (length as a function of time) to reach a steady state after some time. Steady state is the regime where the average filament length does not change with time, and the instantaneous length, $l(t)$ merely fluctuates about the average. In this regime, one can analyze a length-control mechanism and make simple predictions about the steady state length of the structure using mathematical tools from statistical physics. The quantity that we are most interested in is the steady state length distribution of the filament, which we will describe in more detail in the next section. We argue that the specific mathematical form of the length distribution provides a sensitive quantitative lens for viewing cytoskeletal filaments, and complements the more traditional experimental lenses used in microscopy.

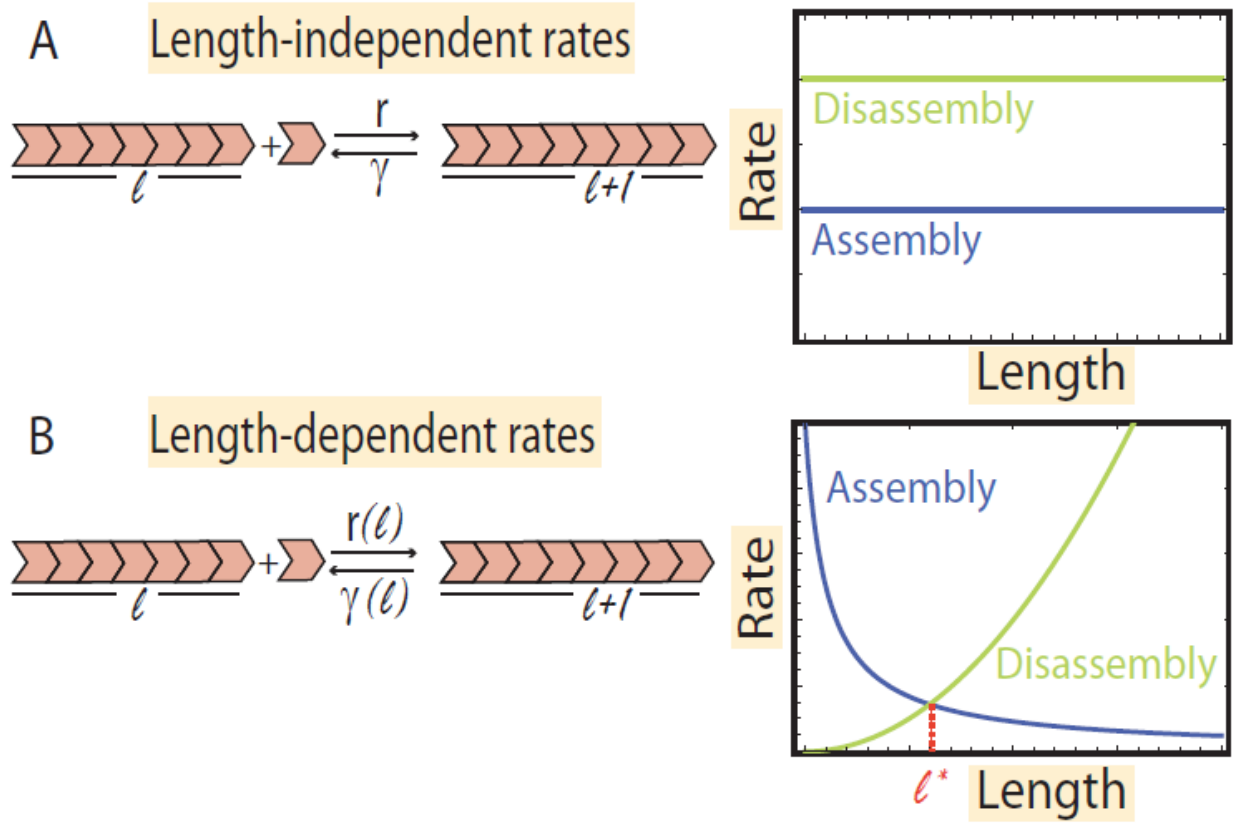

**Figure 2: Length-dependent rates of assembly and disassembly of a dynamic filament are required for length control.** For a dynamical filamentous structure to achieve a well-defined length the rate of addition of monomers must be balanced by the rate of their removal. (A) The graphs for the rate of monomer addition ($r$, blue) and removal ($\gamma$, green) as a function of length do not intersect if the rates are length-independent, and no well-defined length is reached in steady state (Section 2.1). (B) When the rates are length-dependent, the rate curves can intersect resulting in a filament with a well-defined steady-state length $l^*$.

While studying an underlying mechanism of length control it is important to consider that control of the size of a structure must involve feedback between the rates of addition and removal of the monomers and the size of the structure itself. In other words, if the rates of addition and removal of the monomers are length independent, then these rates cannot balance each other at a particular length and will not result in a structure with a peaked distribution of lengths (**Figure 2A**). However if the rates are length-dependent then they can in principle balance each other and thus lead to a well defined average filament length, with small fluctuations in length around the mean (**Figure 2B**). Furthermore, for a stable steady state to be achieved this feedback must be negative, i.e., a longer filament should have a slower assembly rate or a faster disassembly rate.

In this review, we do not discuss explicitly the nucleotide state of the monomers within the filament, other than the effect it may have on proteins that are involved in length regulation (e.g., the preference of the severing protein cofilin for ADP-actin over ATP-actin). In vitro, the nucleotide state can have a significant effect on length control of a filament by modulating its assembly and disassembly dynamics. In fact, a recent study showed that monomer hydrolysis can generate a length-dependent disassembly rate, and hence lead to peaked length distributions for filaments (12).

Herein we focus on the role that actin- and microtubule-associated proteins play in the control of filament length. Experimental evidence suggests that this is a defining feature of the mechanisms at play in cells, namely that proteins are used to sense size and to provide length-dependent feedback cues to a growing structure. For example, motor proteins like Kip3 disassemble microtubules in a length dependent way (49, 50) and cofilin proteins sever actin filaments in a length-dependent manner (1, 41). In this review, we discuss a number of such mechanisms that cells might use to control length of filamentous structures and classify them in two broad categories: length control by assembly, and length control by disassembly, depending on which process is the target of length-dependent feedback. For each mechanism we use the length distribution of filament as its fingerprint, which distinguishes it from other length-control mechanisms.

*1.3 Master equation for filament length*

At the level of abstraction advocated in this review, there is a simple unifying framework that allows us to examine all of the different length control mechanisms. The key quantity that we compute is $P(l, t)$, which is the probability that the filament has length $l$ (measured in units of monomers) at time $t$.

In order to compute $P(l, t)$, we consider a simple model of a single filament exchanging subunits with a free pool of monomers. Note that in this review we consider the assembly of filaments only from fixed nucleation/assembly centers, which are typically defined in the cell by localized proteins (e.g., formins) or protein complexes (e.g., centrosomes).

To make precise the quantity $P(l, t)$ it is useful to consider a population of $N$ filaments where the $i^{th}$ filament has a length $l_i$ at time $t$. We are interested in identifying the frequency (or probability) at which filaments of different lengths will be found in the population. The probability that a filament of length $l$ is found is given by $P(l, t) = \frac{n_l}{N}$, where $n_l$ is the number of filaments of length $l$ at time $t$ (**Figure 3A**).

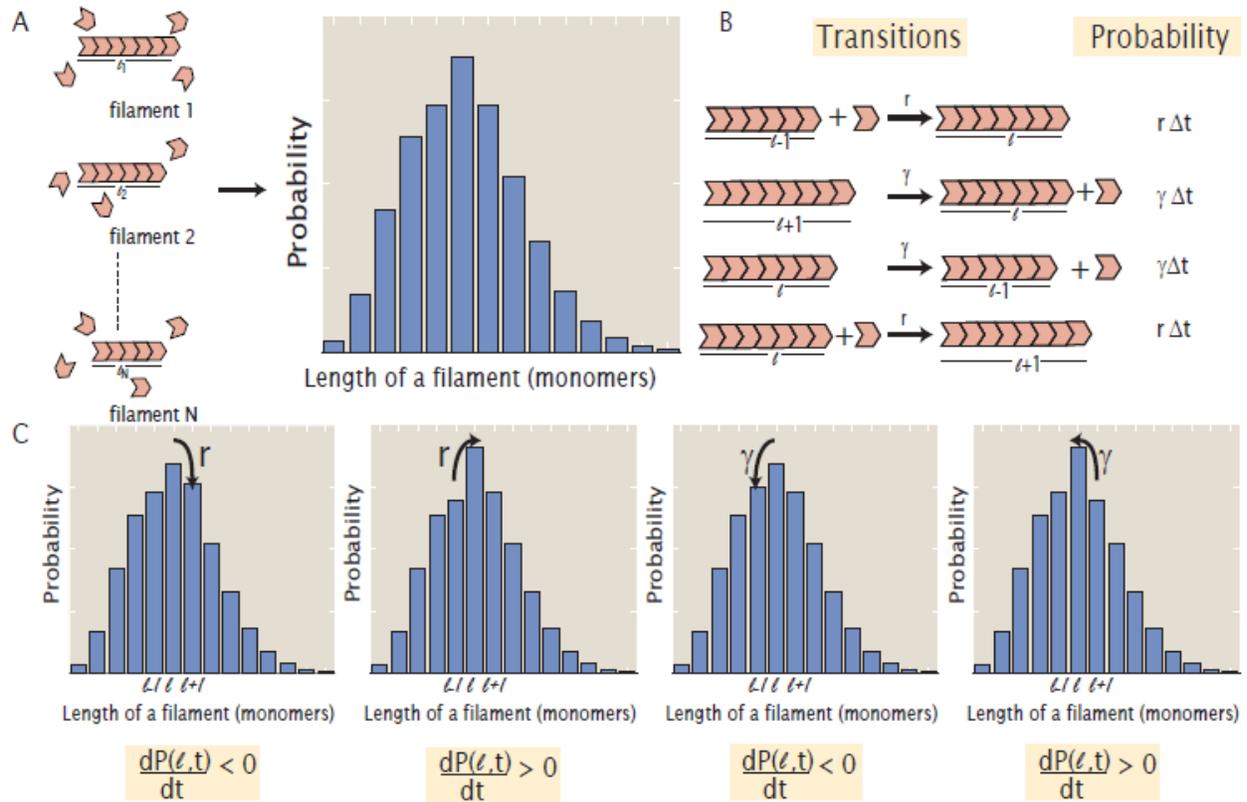

**Figure 3: Master equation for dynamics of filament assembly.** (A) $P(l,t)$ is the frequency of occurrence of filaments of length $l$ in an ensemble of dynamical filaments at time $t$. (B) List of all transitions that lead to a change in the probability of a filament having a length $l$. A filament can either grow starting as a filament of length $l-1$ by adding a subunit with a rate $r$ or shrink by starting as a filament of length $l+1$ by losing a subunit with a rate $\gamma$. These transitions will increase $P(l,t)$ by $r\Delta t$ and $\gamma\Delta t$ respectively. Alternatively, a filament of length $l$ can either shorten to a filament of length $l-1$ with a rate $\gamma$ or add a subunit and become a filament with length $l+1$ with a rate $r$. These transitions will decrease $P(l,t)$ by $\gamma\Delta t$ and $r\Delta t$, respectively. (C) Changes in the probability distribution due to different molecular processes, described in (B).

Next we are interested in how the probability that a filament has a specific length changes with time. We consider a filament that grows by the addition of a subunit at a rate $r$ and shrinks by the loss of a subunit with a rate $\gamma$. In **Figure 3B**, we illustrate all the possible transitions that involve a filament of length $l$ and their contribution to the change in probability $P(l, t)$. In order to produce a filament of length $l$ at time $t + \Delta t$, a filament at time $t$ can either grow from a filament of length $l - 1$ by adding a subunit with a rate $r$, or it can shrink from a filament of length $l + 1$ by losing a subunit with a rate $\gamma$. Note that these terms will increase the probability $P(l, t)$. Alternatively, a filament of length $l$ can either shrink to a filament of length $l - 1$ with a rate $\gamma$ or add a subunit and become a filament of length $l + 1$ with a rate $r$. These terms will reduce the probability $P(l, t)$. Since all these transitions occur during the time interval $\Delta t$, the contribution of any of these to the probability $P(l, t)$ will be given by the product of their corresponding rate with time $\Delta t$. Translating these words into mathematics leads to the equation

$$P(l, t + \Delta t) = P(l, t) + r\Delta t\, P(l - 1, t) - r\Delta t\, P(l, t) + \gamma\Delta t\, P(l + 1, t) - \gamma\Delta t\, P(l, t)\,, \quad (1)$$

where $P(l, t + \Delta t)$ is the probability that the filament has a length $l$ at time $t + \Delta t$. Moving $P(l, t)$ to the left hand side of Equation 1 and then dividing both sides by $\Delta t$, we obtain

$$\frac{P(l, t + \Delta t) - P(l, t)}{\Delta t} = rP(l - 1, t) - r\, P(l, t) + \gamma\, P(l + 1, t) - \gamma\, P(l, t)\,. \quad (2)$$

For very short time intervals $\Delta t$, the left hand side of the equation reduces to the time derivative of the probability and we are left with a *master equation*, namely

$$\frac{dP(l, t)}{dt} = rP(l - 1, t) - r\, P(l, t) + \gamma\, P(l + 1, t) - \gamma\, P(l, t)\,. \quad (3)$$

Note that in Equation 3, the first and third terms represent the inflow of probability to state $l$ while the second and fourth terms represent the outflow of probability from the same state. This is shown pictorially in **Figure 3C**. Note that the rates $r$ and $\gamma$ can themselves depend on $l$, the length of the filament, and this will lead us to a number of interesting and distinct scenarios. Indeed, one of the goals of this review is to show how such ideas can be applied to a number of different cytoskeletal structures and thereby provide a unifying framework by which distinct models of length control can all be viewed from the same perspective.

*1.4 Solution of the master equation*

For each of the different length control mechanisms we describe later, the quantity that we are interested in is the steady state probability distribution of filament lengths. In order to solve the corresponding master equations for each mechanism, we use the scheme outlined next.

At steady state, the probability function no longer changes in time, i.e., $\frac{dP(l,t)}{dt} = 0$. Hence Equation 3 becomes

$$r\, P(l-1) + \gamma\, P(l+1) = (r+\gamma)P(l). \quad (4)$$

Note that $r$ and $\gamma$ can depend on length, i.e., they are described by functions $r(l)$ and $\gamma(l)$, and as we will see in the forthcoming sections this has many interesting repercussions. Consider Equation 4 for $l = 0$, which reads,

$$r(-1)\, P(-1) + \gamma(1)\, P(1) = \big(r(0) + \gamma(0)\big)P(0). \quad (5)$$

Since a filament of negative length is meaningless we impose the condition that $P(l) = 0$ if $l$ is a negative integer; similarly, there can be no removal of monomers from the $l = 0$ state. Hence

Equation 5 becomes

$$P(1) = \frac{r(0)}{\gamma(1)} P(0), \quad (6)$$

so we can compute $P(1)$ in terms of $P(0)$. Since we are considering the growth of a filament from a nucleating center, $P(0)$ is interpreted to mean that no monomers are attached to the nucleating center.

Next, we consider Equation 4 for $l = 1$, namely

$$r(0) P(0) + \gamma(2) P(2) = (r(1) + \gamma(2)) P(1). \quad (7)$$

Substituting Equation 6 in Equation 7, we find an expression for $P(2)$ in terms of $P(0)$, namely,

$$P(2) = \frac{\left[(r(1) + \gamma(2))\left(\frac{r(0)}{\gamma(1)}\right) - r(0)\right]}{\gamma(2)} P(0). \quad (8)$$

Proceeding in this way, we produce a general expression for $P(l)$ in terms of $P(0)$, namely

$$P(l) = f(r(l), r(l-1), \ldots r(0), \gamma(l), \gamma(l-1), \ldots \gamma(0)) P(0). \quad (9)$$

In section 2, we will see that the function $f(r(l)..,\gamma(l)..)$ will be considerably simpler once we put in the actual length dependence specific to a given length control mechanism.

Interestingly, this recipe is equivalent to solving for the steady state probability distribution by balancing assembly from the state $l$ and disassembly from the state $l+1$, namely

$$r(l) P(l) = \gamma(l+1) P(l+1), \quad (10)$$

a condition called detailed balance. Then by invoking recursion, we are able to obtain an expression for $P(l)$ in terms of $P(0)$, which is exactly the one obtained in Equation 9.

All that is left is to compute $P(0)$. This is done using the normalization condition $\sum_{l=0}^{\infty} P(l)=1$, which states that the total probability of having any number of monomers in the filament (including zero) is one. In section 2, we use this simple recursive scheme to solve master equations for a number of different length control mechanisms, where the only change from one mechanism to the next is the length dependence of rates.

| Mechanism | Steady state probability distribution | Mean |
|---|---|---|
| A. Finite monomer pool | $\left(\dfrac{r'}{\gamma}\right)^{l-N} \dfrac{N!}{(N-l)!} \dfrac{e^{-\frac{\gamma}{r'}}}{\Gamma\left(N+1,\frac{\gamma}{r'}\right)}$ | $N - \dfrac{\gamma}{r'}$ |
| B. Diffusing dampers | $\left(1-\dfrac{\bar{r}}{\gamma}\right)\left(\dfrac{\bar{r}}{\gamma}\right)^{l}, \quad \bar{r} = r\left(\dfrac{k_{off}}{k_{on}+k_{off}}\right)$ | $\dfrac{\bar{r}}{\gamma - \bar{r}}$ |
| C. Transported dampers | $\dfrac{\left(\dfrac{rk_{off}}{\gamma w}\right)^{\frac{k_{off}}{w}+l-1} e^{-\left(\frac{k_{off} r}{\gamma w}\right)} \Gamma\left(\frac{k_{off}}{w}-1\right)}{\Gamma\left(\frac{k_{off}}{w}+l\right)} \left(\Gamma\left(\dfrac{k_{off}}{w}-1\right) - \Gamma\left(\dfrac{k_{off}}{w}-1,\dfrac{k_{off} r}{\gamma w}\right)\right)^{-1}$ | $\dfrac{k_{off}}{w}\left(\dfrac{r}{\gamma} - 1\right)$ |
| D. Active transport of free monomers | $\left(\dfrac{r'}{\gamma}\right)^{l-1} \dfrac{e^{-\left(\frac{r'}{\gamma}\right)}}{(l-1)!}$ | $\dfrac{r'}{\gamma} + 1$ |
| E. Depolymerizers | $\left(\dfrac{r}{\gamma'}\right)^{l} \dfrac{e^{-\left(\frac{r}{\gamma'}\right)}}{l!}$ | $\dfrac{r}{\gamma'}$ |
| F. Severing | $\dfrac{l\,s\,r^{l-1}}{(r+s)(r+2s)(r+3s)\ldots(r+ls)}$ | $e^{\frac{r}{s}}\left(\dfrac{r}{s}\right)^{1-\frac{r}{s}}\left(\Gamma\left(\dfrac{r}{s}\right) - \Gamma\left(\dfrac{r}{s},\dfrac{r}{s}\right)\right)$ |

**Table 1: Length-control mechanisms in cells.** Formulas for the steady-state probability distributions, and for their means, for different length control mechanisms observed in cells. Each mechanism is discussed in detail in the section indicated in the first column. Note that cases B and C are computed in the limit of fast switching rates when compared to assembly and disassembly rates (see Sec. 2.2.2). If that limit is not met the distributions can be computed numerically.

The key idea of this review is to use length distributions as a way to characterize and distinguish between different length-control mechanisms. In order to motivate the reader, we cut to the chase and present all of our results in Table 1, where we list the different mechanisms that we will discuss in this review. Along with the corresponding closed-form steady state length distributions for each mechanism, we also list equations for the mean of the length distribution as a function of parameters related to the assembly and disassembly of the filament, specific to each mechanism. By varying any of these specific parameters, one can predict changes in the properties of the length distribution for each mechanism; we expand on this idea in the Discussion. In this review, we discuss several mechanisms, but if one is of particular interest, we encourage the reader to go directly to that section, as indicated in **Table 1**.

## *2. Mechanisms of length control*

### *2.1 Unregulated filament*

We first consider a generic filament, which adds and loses subunits at rates $r$ and $\gamma$, respectively with no length dependence (**Figure 4A**). This filament could be a microtubule, adding or subtracting monomers at its plus end or an actin filament adding subunits primarily at its barbed/ plus end and losing them from its pointed/ minus end.

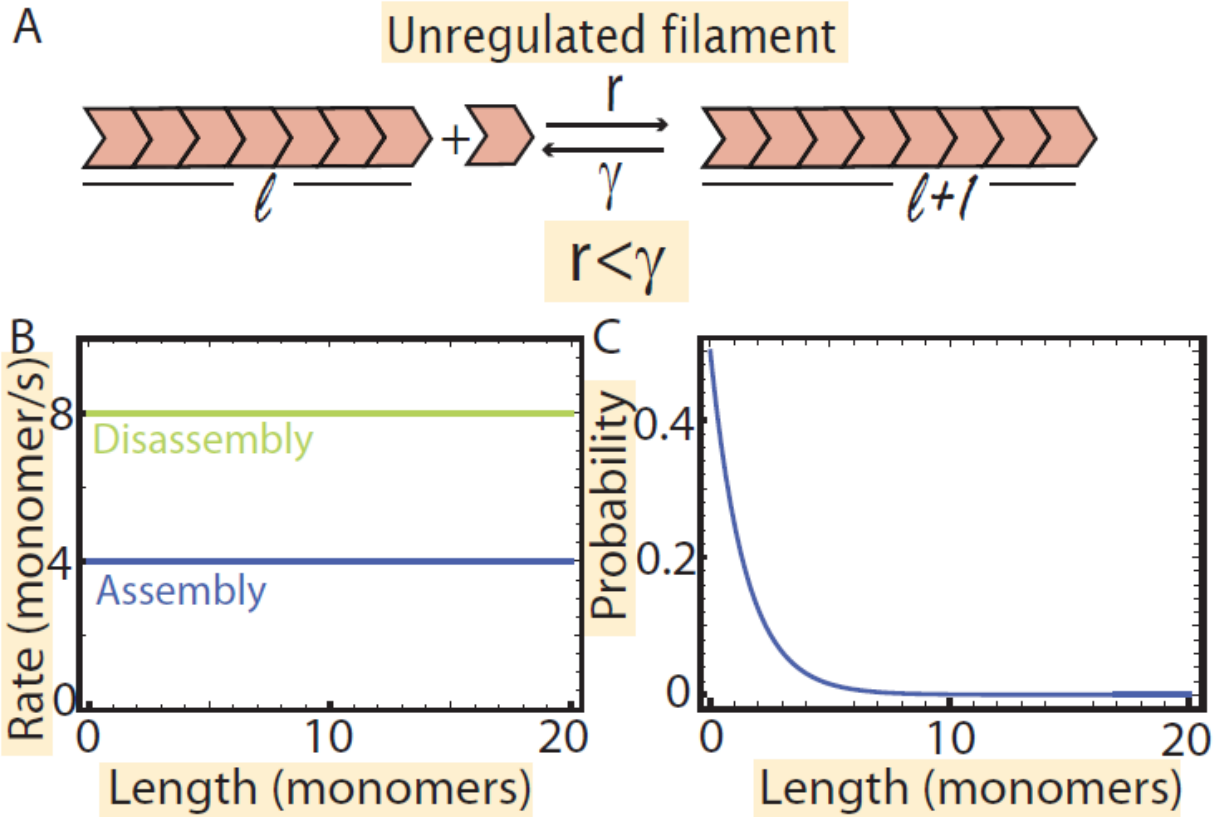

**Figure 4: Unregulated filament.** (A) Schematic of a dynamic filament with length-independent rates of addition ($r$) and removal ($\gamma$) of monomers. (B) If the rate of removal is higher than the rate of addition of monomers the distribution of filament lengths in steady state is exponential (C) The values of the rates used to generate the plot are $r = 4\ s^{-1}$ and $\gamma = 8\ s^{-1}$, are not obtained from experiment and were chosen just for purposes of illustration.

*2.1.1 Master equation for unregulated filament*

The master equation for an unregulated filament is given by Equation 3 in Section 1.3, namely

$$\frac{dP(l,t)}{dt} = r\,P(l-1) - rP(l) + \gamma\,P(l+1) - \gamma\,P(l). \qquad (11)$$

Given the two rates, $r$ and $\gamma$, which are both length-independent, we consider two possible scenarios, $r > \gamma$ and $r < \gamma$.

If the rate of subunit addition is larger than the rate of subunit loss ($r > \gamma$), the filament grows indefinitely (while subunits are available), and steady state is not reached. In fact, the average filament length $\langle l(t) \rangle = (r - \gamma)t$ at late times (See SI 1.1 for a detailed mathematical analysis), implying that the filament will grow indefinitely with time. Note that here we are considering an infinite pool of monomers. We consider the effect of a finite monomer pool in Section 2.2.1.

In the regime where the rate of addition of monomers is less than the rate of removal of monomers ($r < \gamma$), the filament shrinks more than grow and can always shrink down to zero, hence there is a steady state and $P(0) \neq 0$. As we will see in the next calculation, this process actually leads to an exponential distribution for $P(l)$. The master equation for this is identical to the one described in Equation 11, but in this can be exactly solved at steady state, where

$$r P(l-1) + \gamma P(l+1) = (r+\gamma)P(l). \quad (12)$$

In order to find $P(l)$, we use the scheme described in section 1.4. At steady state, Equation 11 can be solved recursively to obtain a general expression of the form $P(l) = \left(\frac{r}{\gamma}\right)^l P(0)$ (see SI 1.2 for the detailed calculation). Using the normalization condition, $\sum_{l=0}^{\infty} P(l)=1$, we can obtain an expression for $P(0)$ and have the complete expression for $P(l)$ as $\left(\frac{r}{\gamma}\right)^l \left(1 - \frac{r}{\gamma}\right)$. The average length of this distribution is give by $\frac{r}{\gamma-r}$ and the variance is $\frac{r\gamma}{(\gamma-r)^2}$. Note that this distribution can have a large average length only if the rates of addition and removal of monomers are closely matched. In the absence of a mechanism that leads to such tight balancing of rates, the average length is small.

The key conclusion of this calculation is that both scenarios ($r > \gamma$ and $r < \gamma$) are unable to give a peaked distribution of lengths, which is a signature of a length-control mechanism. In the case of $r > \gamma$, the filament grows indefinitely and has no steady state whereas in the case of $r < \gamma$, the steady state filament length distribution is exponential. In order to obtain a peaked distribution of lengths, either the addition or the removal rate has to be length-dependent.

## 2.2 Length control by assembly

### 2.2.1 Finite subunit pool mechanism

A finite pool of building blocks (actin or tubulin subunits) is a simple mechanism to control filament length. The assembly rate of a filament depends on the concentration of free subunits. In a finite system, as the filament grows, the free subunit concentration decreases, thus leading to a decrease in the assembly rate (17). When the assembly rate equals the disassembly rate, steady state is reached.

In the following calculation we consider the case of a single filament within a pool of subunits. Let $N_t$ be the total number of total subunits in a cell (**Figure 5A**). As the average assembly rate for the filament is proportional to the concentration of free subunits, the first-order rate constant for assembly can be written down as the second order one times $C_f$, the concentration of free subunits in the cell. Initially, when all the subunits are free in the cell, $C_f = \frac{N_t}{V_c}$, where $V_c$ is the volume of the cell, and we can write the assembly rate as $r = r'N_t$. Here $r'$ is the second order rate constant divided by the cell volume. As more and more subunits get incorporated into the filament, the length of the filament increases and consequently the concentration of free subunits decreases. This results in a length-dependent assembly rate, namely is $r(l) = r'(N_t - l)$, where $l$ is the length of the filament (**Figure 5B**). The disassembly rate $\gamma$ of the filament we take to be constant.

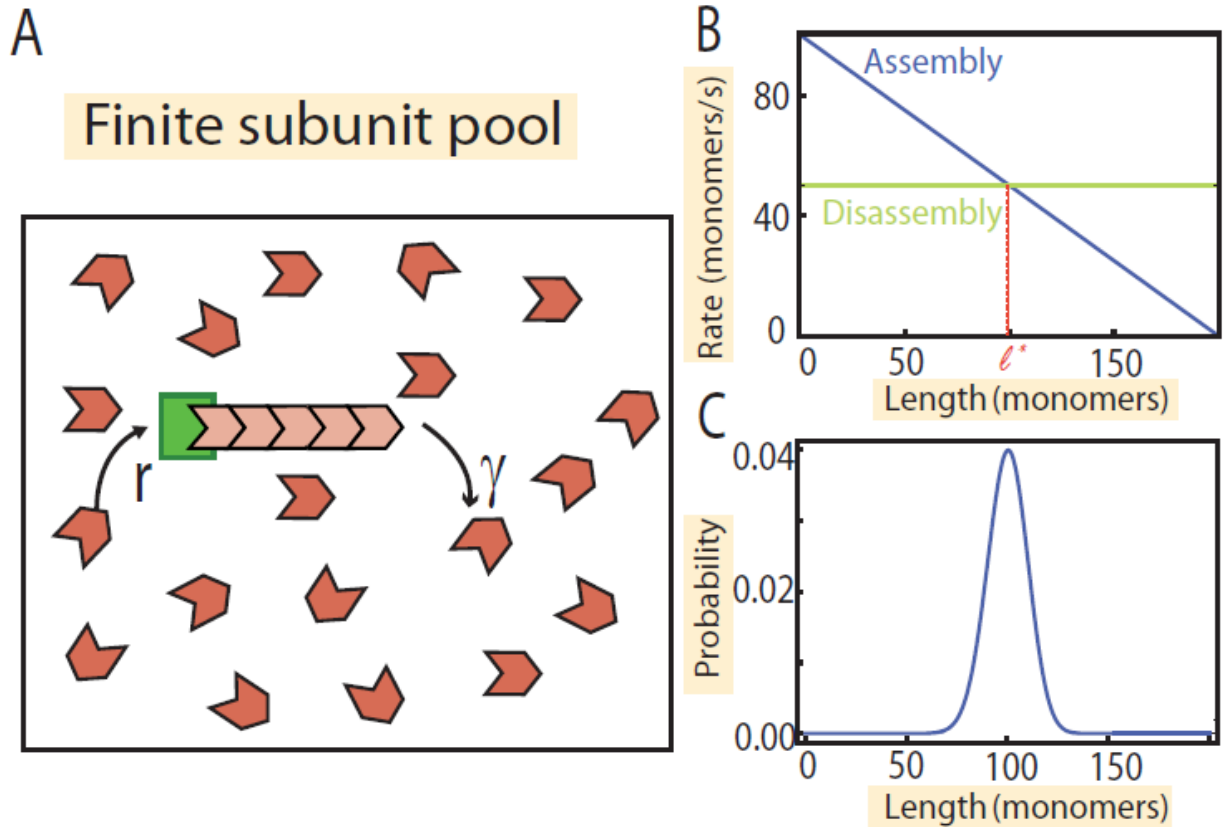

**Figure 5: Finite pool of subunits as a length control mechanism** (A) Schematic of the model. We model a cell as a compartment with one filament (orange) and a fixed total number of subunits ($N_t$, red and orange) within the compartment. The subunits that are not in the filament (red) are freely diffusing in the cell compartment. (B) The assembly rate of the filament ($r = r'(N_t - l)$, blue) decreases linearly with its length, due to the depletion of the free subunit pool. Coupled with a constant disassembly rate ($\gamma$, green) this process can lead to a well defined steady-state length $l^*$. (C) The steady-state length distribution is a peaked function around the mean filament length. For purposes of illustration, we chose the rates $r'$= 0.5/(subunits s), $\gamma$= 5/s and $N_t$=200 subunits.

*2.2.1.1 Master equation for finite subunit pool mechanism*

We can write the master equation for a filament in a finite pool of subunits in the following way,

$$\frac{dP(l,t)}{dt} = r(l-1)P(l-1) - r(l)P(l) + \gamma P(l+1) - \gamma P(l). \quad (13)$$

This equation is similar to Equation 3 derived earlier in Section 1.3, with one key difference. The assembly rate is now length dependent. To summarize, in order to produce a filament of length $l$, the filament can either grow from $l-1$ by adding a subunit with a rate $r(l-1)$ that depends on the length of the filament, or shrink from $l+1$ by subtracting a subunit with a rate $\gamma$. Alternatively the filament of length $l$ can either degrade to $l-1$ with a rate $\gamma$ or add a subunit and acquire length $l+1$ with a length dependent rate $r(l)$.

At steady state, from Equation 13, we get $r(l-1) P(l-1) + \gamma P(l+1) = (r(l) + \gamma)P(l)$. This equation can be solved recursively using the scheme described in Section 1.4 to obtain $P(l)$ in terms of $P(0)$, namely

$$P(l) = \left(\frac{r'}{\gamma}\right)^l \frac{N_t!}{(N_t - l)!} P(0). \quad (14)$$

Using the normalization condition, we find $P(0) = \left(\frac{r'}{\gamma}\right)^{-N_t} \frac{e^{-\frac{\gamma}{r'}}}{\Gamma\left(N_t+1, \frac{\gamma}{r'}\right)}$, where $\Gamma(x, y)$ is the incomplete Gamma function (see SI 2.1 for the detailed calculation). $P(l)$ is a distribution peaked around the mean length (**Figure 5C**).

Next we compute the dependence of the moments of the distribution on the various parameters of the model, namely $N_t$, $r'$ and $\gamma$. These will be specific predictions of this model, and by designing experiments where these parameters are tuned, one can test whether the length of a filament is indeed being controlled by the finite subunit pool mechanism.

The mean of the distribution (Equation 14) is $\langle l \rangle = N_t - \frac{\gamma}{r'}$ (see SI 2.2 for a detailed calculation). Note that this average length can also be computed by equating the average rate of assembly with the rate of disassembly $\gamma$, namely

$$r'\left(N_t - \langle l \rangle\right) = \gamma. \quad (15)$$

The key prediction of this equation is that increasing the free pool of subunits (i.e., increase in $N_t$) increases the average filament length, whereas increasing the disassembly rate (i.e., increase in $\gamma$), which can be achieved by increasing the concentration of disassembly factors in the cell, decreases the average filament length.

The variance of the distribution in steady state is $\langle l^2 \rangle - \langle l \rangle^2 = \frac{\gamma}{r'}$, (See SI 2.3 for a detailed calculation) which interestingly does not depend on the size of the total monomer pool. The functional dependence of the mean and the variance of the length, on the number of free monomers is a signature of the finite-subunit pool mechanism. We discuss this later in section 3, where we discuss ways to characterize and distinguish between different length control mechanisms.

*2.2.1.2 Actin cables in budding yeast cells*

In budding yeast cells, free actin subunits are used by formins to build linear actin structures known as cables, which are used as polarized tracks for intracellular transport (5). Interestingly, in wild type cells the cables rarely grow longer than the diameter of the mother cell compartment. Here, using simple estimates, we assess whether the cable length is controlled by the finite pool mechanism.

For the purposes of an estimate, we assume a simple geometry for the cables, where each cable has an average length $\langle l \rangle$, and it consists of $D$ actin filaments in parallel bundled together. In the presence of a finite monomer pool of actin subunits, the average assembly rate can be estimated as $r'(N_t - N_c D \langle l \rangle)$, where $N_t$ is the total number of actin molecules in the cell (in both filamentous and monomeric forms), $N_c$ is the number of cables, and $r'$ is the assembly rate per free monomer; note that in the absence of cables, when all of the actin molecules are in monomeric form, $r = r'N_t$. At steady state, the average assembly rate is equal to the disassembly rate $\gamma$, which leads to an average cable length $\langle l \rangle = (N_t - \gamma/r')/N_c D$.

The total number of actin molecules in the mother-cell can be estimated by considering the concentration of actin in the cell's cytoplasm, which has been measured by quantitative western blotting (Johnston et al, in press), and multiplying it by the known volume of a budding yeast mother-cell, $N_t = 10 \text{ μM} \times \frac{4\pi}{3}(2.5 \text{ μm})^3 = 4 \times 10^5$ actin proteins. Observations in vivo suggest that the number of cables is roughly $N_c = 10$ and each of them have about $D=4$ filaments. Furthermore, by using the measured in vivo rate of cable assembly, $r = 370 \text{ s}^{-1}$(52) ($r' = r/N_t = 10^{-3} \text{ s}^{-1}$), and an estimated rate of disassembly $\gamma = 45 \text{ s}^{-1}$ (38), we estimate an average cable length of $\langle l \rangle = 30$ μm; we use the conversion 1 μm = 370 monomers. (This estimate purposely does not consider the rate of actin incorporation into another type of structure called actin patches, which are concentrated primarily in the bud.) Interestingly, this estimated average cable length is about a factor of five longer than what is observed in wild-type yeast cells. Furthermore, a formin binding protein (Smy1) was identified recently, which modulates the length of actin cables. Our estimate and this observation are both indicative of the finite subunit pool mechanism not being solely responsible for cable length control in budding yeast (39, 52).

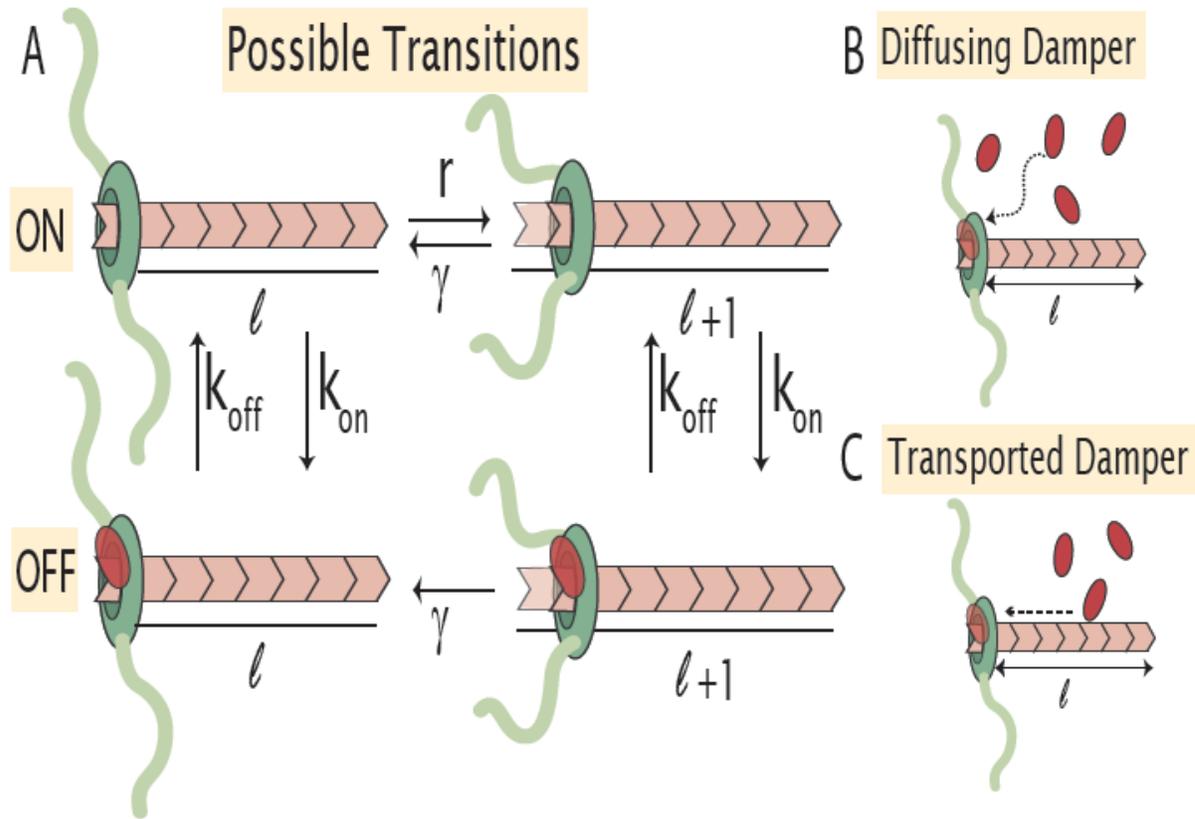

**Figure 6: Inhibition of elongator-driven filament growth by dampers.** (A) An elongator (green) can add a subunit to an existing filament to increase the filament length with a rate $r$. The filament can lose a subunit, and decrease the filament length by one unit, with a rate $\gamma$. However if a damper is associated with the elongator, then a filament either pauses growth or exhibits a reduced growth rate. The damper attaches to the elongator with rate $k_{on}$ and falls off with rate $k_{off}$. The filament can lose a subunit with a rate $\gamma$, even with the damper attached. A damper can reach the elongator by free diffusion (B) or by directed transport on the filament itself (C).

*2.2.2 Elongators and dampers*

Elongators are proteins that associate with the growing end of a filament and increase the rate of its assembly. For example, formin proteins can associate with the barbed end of an actin filament and increase its assembly rate by many fold. In fact they can support actin assembly in vitro at rates up to 55 subunits/(s μM), compared to the rate of 11 subunit/( s μM ) for assembly of actin filaments without formins (28, 42) . Since the assembly rate is much greater than the pointed - end disassembly rate of an actin filament (about 0.25 subunit/s) (42), one way that cells can modulate the rate of assembly by formins is using a molecular 'damper', which is a protein that associates with the elongator to reduce the rate of assemblySmy1, Hof1 and Bud14 are examples of formin dampers in budding yeast cells (7, 19, 20).

Because of the inhibitory activity of a damper, the elongator has two states. It is in the ON state when the damper is not associated with it and in the OFF state when the damper is bound to it (**Figure 6A**). Since the damper can bind and unbind from the elongator, there are rates connecting these two states: $k_{off}$, the rate at which the OFF state transitions to the ON state will depend on the binding affinity of the damper for the elongator, and $k_{on}$, the rate at which the ON state transitions to the OFF state will depend on how the damper is delivered or diffuses to the elongator. Note that in the OFF state, the filament has a reduced assembly rate or may not grow at all.

We consider two ways by which dampers can reach the elongator (**Figure 6B** and **6C**), either by free diffusion or by active transport along the filament, and compute the resulting distribution of filament lengths in each case.

**Diffusing damper**: This mechanism is characterized by two additional parameters, the rate at which the damper reaches the elongator by diffusion ($k_{on}$), and the rate at which the damper dissociates from the elongator ($k_{off}$); **Figures 6A**. The average time the filament spends in the ON state, when the elongator is active and the filament is growing at rate $r$, is $1/k_{on}$, while the average time the filament spends in the OFF state is $1/k_{off}$. If we assume that the rate of growth in the OFF state is zero, the average rate of assembly is $\bar{r} = r\left(\frac{k_{off}}{k_{off}+k_{on}}\right)$, where the factor appearing in parenthesis is the fraction of time that the filament spends in the ON state. Note that the average assembly rate ($\bar{r}$) is less than $r$ as $k_{on}$ and $k_{off}$ are both positive.

The average assembly rate $\bar{r}$ and the disassembly rate ($\gamma$) are filament-length independent. This implies that the predictions of this model are qualitatively similar to the unregulated filament model. Namely, there are two scenarios, one corresponding to $\bar{r} < \gamma$ in which case the filaments will be very short and broadly distributed, and $\bar{r} > \gamma$ which in the presence of a finite monomer pool will lead to a sharply peaked distribution around the mean filament length.

Candidate proteins that act as diffusing dampers in budding yeast are Bud14 molecules which are thought to reach the formins localized at the bud neck by diffusion, and are known to inhibit their polymerizing activity (8, 19). From experiments, the binding of Bud14 to formin is described by a dissociation constant $K_d$= 15 nM (8). If we estimate its rate of dissociation from formin to be similar to that of Smy1(another damper, discussed in detail below) then $k_{off} \approx 1/s$, and therefore $k_{on} \approx \left(\frac{k_{off}}{K_d}\right)$[Bud14] $\approx 0.2 \ 1/s$, assuming a Bud14 concentration of 3 nM in a yeast cell (16). For an actin cable in yeast, $r = 370/s$ and $\gamma = 45/s$ (39, 52), hence we expect this damper to reduce the rate of assembly to $\bar{r} = 310/s$. Since $\bar{r} > \gamma$ still holds, a well defined length will not be the result of this mechanism and another method of length control is required. For example, as in the case of an unregulated filament, the presence of a finite pool of actin monomers will lead to a peaked distribution of filament lengths around the mean in steady state. As was remarked in the case of the unregulated filament, this is unlikely the primary length-control mechanism at play for actin cables in budding yeast, since this reduction in the average assembly rate still leads to an average cable length in steady state about five times the observed length.

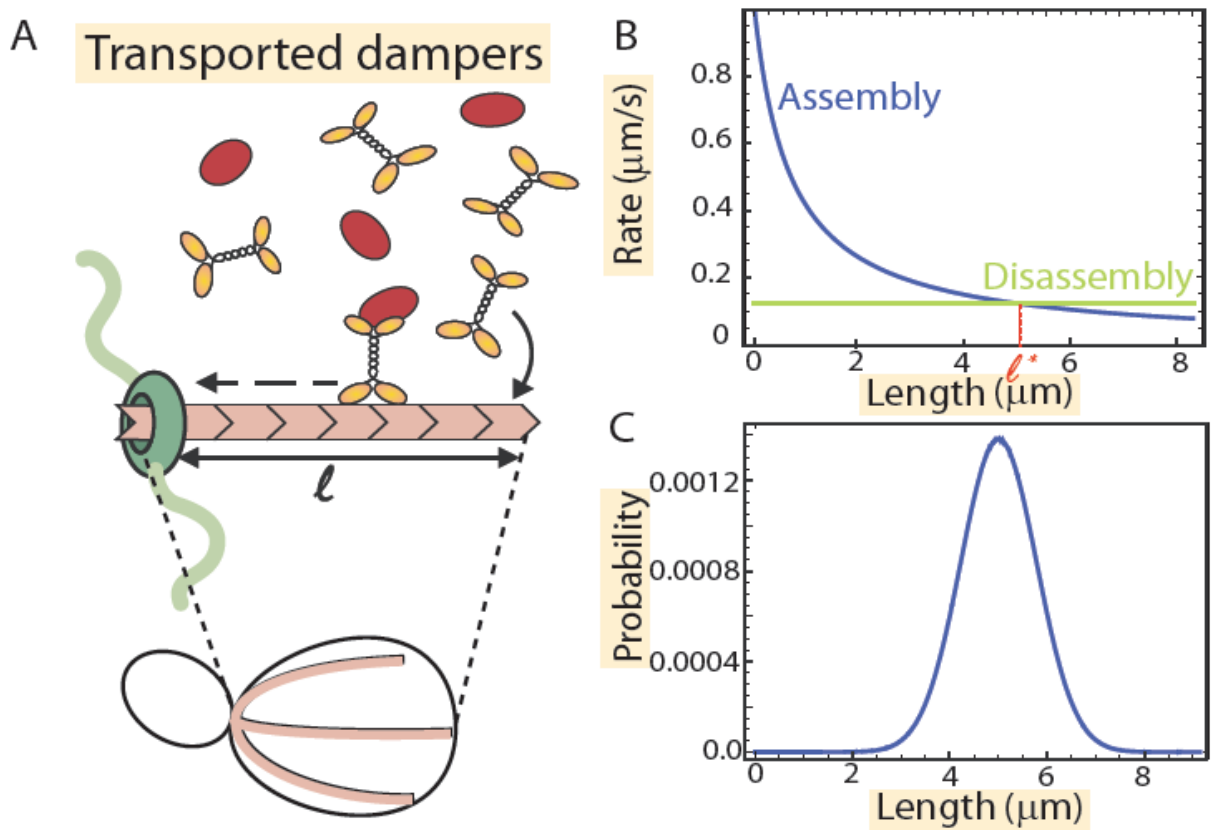

**Figure 7: Active transport of dampers.** (A) Damper proteins (red) can be actively transported on a filament by motors (orange) and delivered to an elongator (green), where they inhibit the activity of the elongator. An example is the case of Smy1 proteins in budding yeast, which are delivered to the formin at the barbed end of the actin cable by myosin motors. Smy1 bound to formin inhibits its polymerization activity. (B) Longer cables will encounter more dampers and hence will provide more inhibition to the elongator as compared to shorter ones, thus setting up a negative feedback. This process leads to a length-dependent assembly rate (blue). In concert with a constant disassembly rate (green) these filament dynamics lead to a peaked length distribution in steady state (C). The parameter values used are $r = 370\ s^{-1}$, $\gamma = 45\ s^{-1}$, $w = 0.004\ s^{-1}$ ($k_{on} = w\ l$) and $k_{off}=1\ s^{-1}$, and they correspond to actin cables in budding yeast (7, 39).

**Actively transported damper**: In contrast to a diffusing damper, a damper molecule can reach the elongator residing on one end of a filament by being actively transported to it by a molecular motor moving along the filament. One example is the budding yeast proteins Smy1 which are actively transported by myosin motor-proteins along actin filaments to the formin Bnr1 anchored at the bud neck of a budding cell (**Figure 7**).

Interestingly, active transport of dampers can give rise to a negative feedback that acts to reduce the assembly rate in a length dependent manner. The filament acts as a landing pad for the myosin motors carrying dampers. Long filaments on average encounter more motors carrying dampers and thereby deliver inhibitory cues at a higher frequency to the elongators at the end of the filament. This can set up a length dependent negative feedback loop whereby longer filaments have a slower average assembly rate.

This mechanism is conceptually similar to the antenna model which has been described in the context of microtubule length control, whereby kinesin motor proteins (Kip3) move directionally on a microtubules and upon reaching its end, stimulate the disassembly of the microtubule (49, 50). A longer microtubule will encounter more such motors and hence will disassemble more rapidly resulting in a well defined steady-state length. We will discuss this mechanism in more detail in section 2.3.1.

*2.2.2.1 Master Equation for the actively transported damper mechanism*

Once again, we use the master equation to describe the dynamics of an individual filament. For a given filament length, we distinguish between two states depending on whether the elongator at its end is inhibited by damper (the OFF state) or free of damper (the ON state). In the ON state, the filament can grow as well as shrink, whereas it only shrinks in the OFF state. Note that for simplicity, here we assume that the rate of filament assembly in the OFF state is zero, but the theory can be simply extended to the more general case of reduced growth in the OFF state. In addition to these processes, the filament can also switch between the two states. Let $k_{on}(l)$ and $k_{off}$ be the rates at which the dampers arrive at the elongator due to the action of motor proteins, and the rate at which a damper falls off from elongator, respectively. Under the assumption of high processivity of myosin+Smy1 complexes, namely, we assume that these complexes do not fall off the filament before reaching its end (we discuss this assumption later), the rate at which Smy1 gets to the formin, $k_{on}(l)$, is equal to the rate of capture of these complexes from solution by the filament. Based on the physics of diffusion to capture, a long filament will capture more complexes compared to a short filament in proportion to their lengths; i. e., $k_{on}(l) = wl$.

The master equation for the filament in the ON state takes into account all the transitions to and from this state, namely

$$\frac{dP_{on}(l,t)}{dt} = r\, P_{on}(l-1) - rP_{on}(l) + \gamma\, P_{on}(l+1) - \gamma\, P_{on}(l) + k_{off}P_{off}(l) - wl\, P_{on}(l). \quad (16)$$

In the ON state, a filament can achieve a length $l$ by adding a monomer (with rate $r$) to a filament of length $l-1$ or by by losing a subunit (with rate $\gamma$) from a filament of length $l+1$. A filament of length $l$ can change length either by losing a subunit or by adding one. A filament in the OFF state can convert to ON state with the rate $k_{off}$ with no change in length. Alternatively, the ON state can turn OFF via binding of the damper with the length dependent rate $wl$, thus reducing $P_{on}$.

Similarly, we write the master equation for the OFF state, in which the filament is not allowed to grow by the presence of the bound damper, namely

$$\frac{dP_{off}(l,t)}{dt} = \gamma\, P_{off}(l+1) - \gamma\, P_{off}(l) - k_{off} P_{off}(l) + wl\, P_{on}(l). \tag{17}$$

We use these master equations to compute the steady-state distribution of filament lengths $P(l) = P_{on}(l) + P_{off}(l)$, where $P_{on}(l)$ and $P_{off}(l)$ are solutions to Equation 16 and 17, when the left-hand sides of these equations are set to zero.

The steady state distribution of filament lengths can be computed exactly using the method of detailed balance in the fast switching regime, i.e., when the rates for switching between the ON and the OFF states ($k_{on}(l)$ and $k_{off}$) are much greater than the rates of assembly/disassembly ($r$ and $\gamma$).

The average time that the filament spends in the ON state, when the elongator is active and the filament is growing at rate $r$, is $1/k_{on}(l)$, while the average time the filament spends in the OFF state is $1/k_{off}$. Since we assume that the rate of growth in the OFF state is zero, the average rate of assembly is $\bar{r}(l) = r\left(\frac{k_{off}}{k_{off}+k_{on}(l)}\right)$, where the factor appearing in parenthesis is the fraction of time that the filament spends in the ON state. We conclude that the average rate of assembly is length dependent and decreases as the length of a filament increases, since $k_{on}(l) = wl$ (**Figure 7B**). Furthermore, the average rate of assembly depends on the concentration of damper since $w$ is proportional to the damper concentration. Also, $k_{off}$ is proportional to the dissociation constant of the binding reaction between the damper and the elongator. Since both parameters can in principle be tuned in experiments, how the distribution of filament changes with these parameters is then an excellent target for a quantitative experimental test of this length control mechanism.

Using the detailed balance condition $P(l)\bar{r}(l) = \gamma P(l+1)$ we obtain

$$P(l)\, r\, \frac{k_{off}}{k_{off} + wl} = \gamma\, P(l+1). \qquad (18)$$

Equation 27 can be recursively solved using the scheme described in Section 1.4 to obtain

$$P(l) = \left(\frac{r}{\gamma}\right)^l \prod_{i=0}^{l-1} \left(\frac{k_{off}}{k_{off} + i\, w}\right) P(0). \qquad (19)$$

We use the normalization condition for $P(l)$ to obtain $P(0)$, which then gives us an analytic formula for the length distribution

$$P(l) = \left(\frac{r}{d}\right)^l \frac{\left(k_{off}/w\right)^{l-1}}{\left(\frac{\Gamma\left(\frac{k_{off}}{w} + l\right)}{\Gamma(l-1)}\right)} \left(\frac{e^{\frac{k_{off}\, r}{\gamma w}} k_{off}\, r (k_{off} - w) \left(\frac{k_{off}\, r}{\gamma w}\right)^{-\left(\frac{k_{off}}{w}\right)} \left(\Gamma\left[\frac{k_{off} - w}{w}\right] - \Gamma\left[-1 + \frac{k_{off}}{w}, \frac{k_{off}\, r}{\gamma w}\right]\right)}{\gamma w^2}\right)^{-1}, \qquad (20)$$

where $\Gamma(x)$ is the Gamma function (See SI 3.1 for detailed calculations). In the regime where the rates of switching are comparable to the rates of assembly and disassembly we can use numerical simulations to solve the master equations from Equation 16 and 17 (39).

From the expression for the average rate of assembly, we can compute the steady-state average filament length by equating it with the rate of disassembly $\gamma$, namely

$$\langle l \rangle = \frac{k_{off}}{w}\left(\frac{r}{\gamma} - 1\right). \qquad (21)$$

The key prediction of this equation is that increasing the damper concentration (i.e., increase in $w$) reduces the average filament length, whereas weakening the elongator-binding affinity of damper (i.e., increase in $k_{off}$) increases the average filament length. This result for the mean length is always correct even when the fast filament assembly limit is not obtained (39).

*2.2.2.3 Key assumptions and further comments*

In this mechanism, it is assumed that the transport of dampers is processive and that each damper that comes on to the filament is able to reach the elongator. This may not be true in general but in the case of budding yeast this is a reasonable assumption in wild type cells, as Smy1-GFP was directly observed to be trafficked by the myosin motor and delivered, uninterrupted, to the formin (7). Smy1 is on vesicles, which may have multiple myosin motors attached to them, and hence processivity does not seem to be an issue. This also validates the assumption in our model that delivery of Smy1 is uninterrupted by motor detachment.

Our calculation assumed that the rate at which damper gets to the elongator is greater than the assembly rate of the filament. If that is not the case, the damper will never be able to catch up to the elongator and hence the length control mechanism will not be in effect. In the case of a wild type budding yeast cell, the observed anterograde transport rate of vesicles toward the bud neck is 3 µm/s (7), which, given a retrograde elongation rate of cables of 0.5-1 µm/s (52), suggests a myosin motor speed of about 3.5-4 µm/s. Hence the assumption that the rate of transport of Smy1 toward the formin is greater than the rate of cable elongation is reasonable.

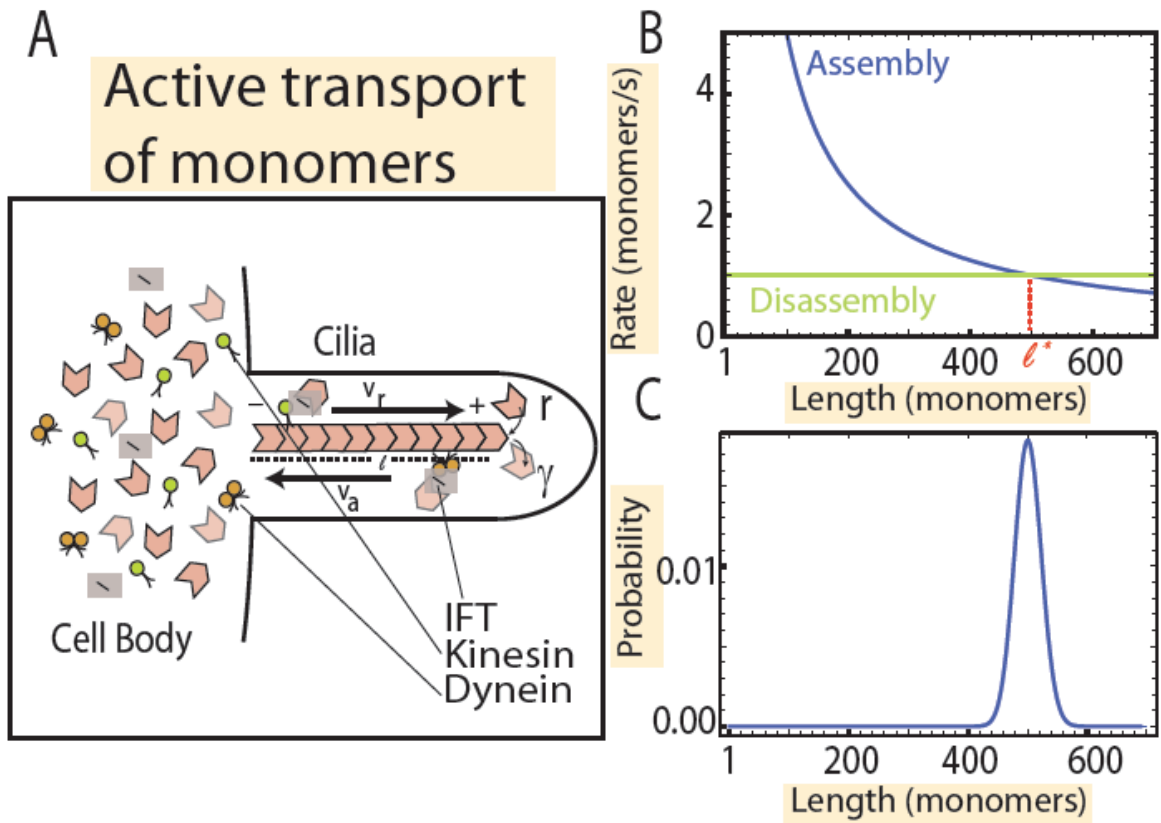

**Figure 8: Active transport of monomers.** (A) Cilia grow and shrink by adding and removing tubulin dimers at the tips of the cilia, which is also the plus end of the microtubules. A fixed number of IFT are loaded into each cilium, and these particles are carried by kinesin motor proteins (green) to the cilia tips, delivering components required for the assembly of the microtubule. They also carry the removed tubulin from the tip with the help of dynein proteins (orange). (B) Shorter microtubules get the building components faster than longer microtubules, hence the assembly rate (blue) is length dependent. Since the disassembly rate (green) is constant, this process can lead to a steady state microtubule length $l^*$. The steady state filament length distribution depends on the assembly and disassembly rate. (C) In the regime, $r' > \gamma$, where the disassembly rate is less than the assembly rate, the distribution is peaked around a The parameters used are $r' = 500\ s^{-1}, \gamma = 1\ s^{-1}$, and were chosen for illustration purposes.

*2.2.3 Active transport of monomers*

In Section 2.2.2, we described a method of length control of actin filaments that relied on active transport with the help of motor proteins. In this Section, we will describe another method of length control that is thought to be active in microtubule-based structures known as cilia(34, 35), and which again relies on active transport, in this case by kinesin and dynein motor proteins.

Cilia are dynamic organelles composed of microtubules. In *Chlamydomonas*, these organelles grow in pairs, usually to around 10 microns in length, and this model organism has long been used to study size control problems. From experiments on cilia, we know that the growth rate of these organelles tapers off with length of cilia but the rate of disassembly of cilia is independent of length (34). It is believed that these two processes balance each other at a steady state length.

The microtubules in cilia are constantly assembled and disassembled by the addition and removal of tubulins that happens primarily at the tip of cilia, which is also the plus end of microtubules (**Figure 8A**) (35). This begs the question as to how the tubulin dimers from the base are reaching the plus end or the tip of the cilia. A simple estimate tells us that by free diffusion a tubulin dimer will take ~ 10 s to traverse the 10 $\mu m$ length of the cillium. This process is rather slow compared to the growth rate of the microtubules where several tens of tubulins are added to the end of the microtubule per sec. As it turns out, active transport of tubulins makes sure that the dimers reach the plus end in time.

The process by which all of the components required for assembly and disassembly are carried up or down the length of the microtubule is known as Intraflagellar Transport (IFT). From experiments with fluorescently labeled antibodies that recognize the IFT proteins, we know that cells load a fixed number of IFT particles (transporters) onto cilia; the amount is independent of the starting cilia length (35). IFT particles walk on the microtubule with the help of kinesin motor proteins towards the tip, and dynein motor proteins away from the tip (See **Figure 8A**). As length increases, the time it takes to transport the building blocks necessary for cilia growth to the tip increases, thus giving rise to a *length dependent* assembly rate. In other words, the longer the length of the microtubules, the more time it would take for the IFT particles to make it to the tip. As a result, longer microtubules grow slowly and smaller microtubules grow quickly. IFT particles also take away the removed components from the tip back to the cell body. In experiments, the rate of disassembly rate of cilia was observed to be independent of length, suggesting that the slow step is detachment of tubulin from the tip and not its transport by dynein (35).

Taking into account the results of these experiments, a simple model for the assembly of cilia was recently proposed (35). If $N$ IFT particles move at a constant rate $v$, then the rate of tubulin delivery is given $r(l) = \frac{r'}{l}$, where $r' = \alpha N v$ and $\alpha$ is proportionality constant. The rate of disassembly $\gamma$ is constant (**Figure 8B**).

*2.2.3.1 Master equation for the active transport of monomers mechanism*

The master equations for the assembly process described above can be written again by taking into account all the molecular-scale transitions that lead to a change in length,

$$\frac{dP(l,t)}{dt} = \frac{r'}{l-1}P(l-1,t) - \frac{r'}{l}P(l,t) + \gamma P(l+1,t) - \gamma P(l,t). \quad (22)$$

This equation is similar to Equation 3 derived in Section 1.3, with the length-dependent assembly rate $r(l) = \frac{r'}{l}$ and the disassembly rate $\gamma$, which is length-independent. In steady state, $\frac{dP(l,t)}{dt} = 0$, and using detailed balance, we obtain $\frac{r'}{l}P(l) = \gamma P(l+1)$. Using recursion scheme described in Section 1.4, we obtain $P(l) = \frac{1}{(l-1)!}\left(\frac{r'}{\gamma}\right)^{l-1} P(1)$; see SI 4.1 for detailed calculations. Using normalization condition for probabilities, $\sum_{l=1}^{\infty} P(l) = 1$ we find $P(1) = e^{-\frac{r'}{\gamma}}$. As a result we have

$$P(l) = \frac{e^{-\frac{r'}{\gamma}}}{(l-1)!}\left(\frac{r'}{\gamma}\right)^{l-1}. \quad (23)$$

Note that in this calculation, we have set $P(0) = 0$ since for this mechanism of active transport to work, we need to start with a filament of non-zero length. In the regime where $r' > \gamma$, the steady state distribution is peaked at lengths that correspond to a large number of monomers (**Figure 8 C**). In the other regime, $r' < \gamma$, the resulting distribution will be exponential (similar to unregulated filament) and hence this mechanism cannot be length controlling.

The distribution in Equation 23 has mean $\frac{r'}{\gamma} + 1$ and variance $\frac{r'}{\gamma}$ respectively. The expression for mean tells us that a cilium with higher assembly rate (or a smaller disassembly rate) will end up having a longer steady state length.

*2.2.3.2 Key assumptions and further comments*

We have made of number of assumption in this calculation, which need to be acknowledged. First, we assumed that the motor proteins are highly processive and that all of them carry IFT particles with cilia building blocks. Second, the model assumes that the number of IFT particles is conserved, i.e. the number of IFT particles moving towards the tip of the cilia is equal to the number on the way back, which may not be the case.

A recent study carefully measured all the rates associated with the growth of cilia in *Chlamydomonas* (35). The rate of assembly of the cilia is given by $r(l) = \frac{\alpha N v}{2l}$ (35). By direct measurements of the motion of IFT particles, $v = 2.5 \frac{\mu m}{s}$, $N = 10$ (based on DIC images), $\gamma = 0.011 \frac{\mu m}{s}$. By using these values the proportionality constant, $\alpha \sim 0.01$ can be estimated for a cilia taking into account its measured steady-state length $\langle l \rangle \sim 10 \ um$. This value can also be independently obtained from measurements of the assembly rate and how it changes with the length of. Given this value of $\alpha$, we can predict the change in cillia length when $N, v$ or $\gamma$ are changed. Also, that measured disassembly rate being much slower than the rate of kinesin/dynein transport is consistent with the observation that disassembly is not length dependent.

## 2.3 Length control by disassembly

### 2.3.1 Depolymerizers

Kinesin motors such as Kip3 and Kif19 disassemble tubulin dimers from the plus end of microtubules(40, 49, 50). The microtubule acts as a landing pad for these motor proteins. Once on the microtubule, these proteins walk towards the plus end, where they stimulate microtubule disassembly (**Figure 9**). A longer microtubule will have a larger number of these motor proteins arriving at the plus end of the microtubule per unit time, leading to a larger disassembly rate for the microtubule. In other words, the removal rate of the subunits is length dependent (49, 50).

Next, we mathematically derive a relationship between the disassembly rate of a filament and the length of the filament. To that end, let us consider a microtubule as a filament with $l$ subunits (**Figure 9A**). Motors arrive on the microtubule by diffusion with the rate $k_{on}$ $(1/s)$. They move on the microtubule with a rate $v$ (motors/ s), this rate is set by the step length of the motor protein. Once the motor reaches the microtubule end, they fall off, taking a tubulin dimer with them. From the physics of diffusion to capture, as in the case of directed dampers discussed in Section 2.2.2, we conclude that the number of kinesins attaching per unit time is proportional to length of the microtubule $l$. Therefore, assuming no kinesins fall off before reaching the end, the number that reach the plus end of the microtubule and subsequently fall off taking tubulin away with them, is proportional to $l$. Hence the disassembly rate of the microtubule $\gamma(l) = \gamma' l$ where $\gamma'$ is the proportionality constant (See SI 5.1 for a detailed calculation).

Now we consider a microtubule as a single filament growing with a rate $r$ and shrinking with a disassembly rate $\gamma' l$ (**Figure 9B**), and use the master equation to analyze this length control mechanism.

*2.3.1.1 Master equations for depolymerizers*

The master equation describing the evolution of $P(l, t)$ that takes into account the processes of monomer addition and subtraction is

$$\frac{dP(l,t)}{dt} = r\, P(l-1) - rP(l) + \gamma'(l+1)\, P(l+1) - \gamma' l\, P(l). \quad (24)$$

Equation 24 is similar to Equation 3 derived in Section 1.3 with the only difference that now the disassembly rate is length dependent, i.e. $\gamma(l) = \gamma' l$. By using the recursion scheme described in Section 1.4, Equation 36 can be solved at steady state to obtain $P(l) = \left(\frac{r}{\gamma'}\right)^l \frac{1}{l!} P(0)$; see SI 5.2 for detailed calculations. Using the normalization relation, $\sum_{i=0}^{\infty} P(i) = 1$, we obtain $P(0) = e^{-\left(\frac{r}{\gamma'}\right)}$. Hence $P(l)$ is a Poisson distribution with a mean $\left(\frac{r}{\gamma'}\right)$. When $\gamma' > r$, i.e., rate of growth is smaller than the disassembly rate, the uncertainty in the length (i.e., the standard deviation of the distribution) is greater than the mean, where as in the regime of $r > \gamma'$, it is less, and in the limit $r \gg \gamma'$ the distribution of filaments in sharply peaked around the mean (**Figure 9**).

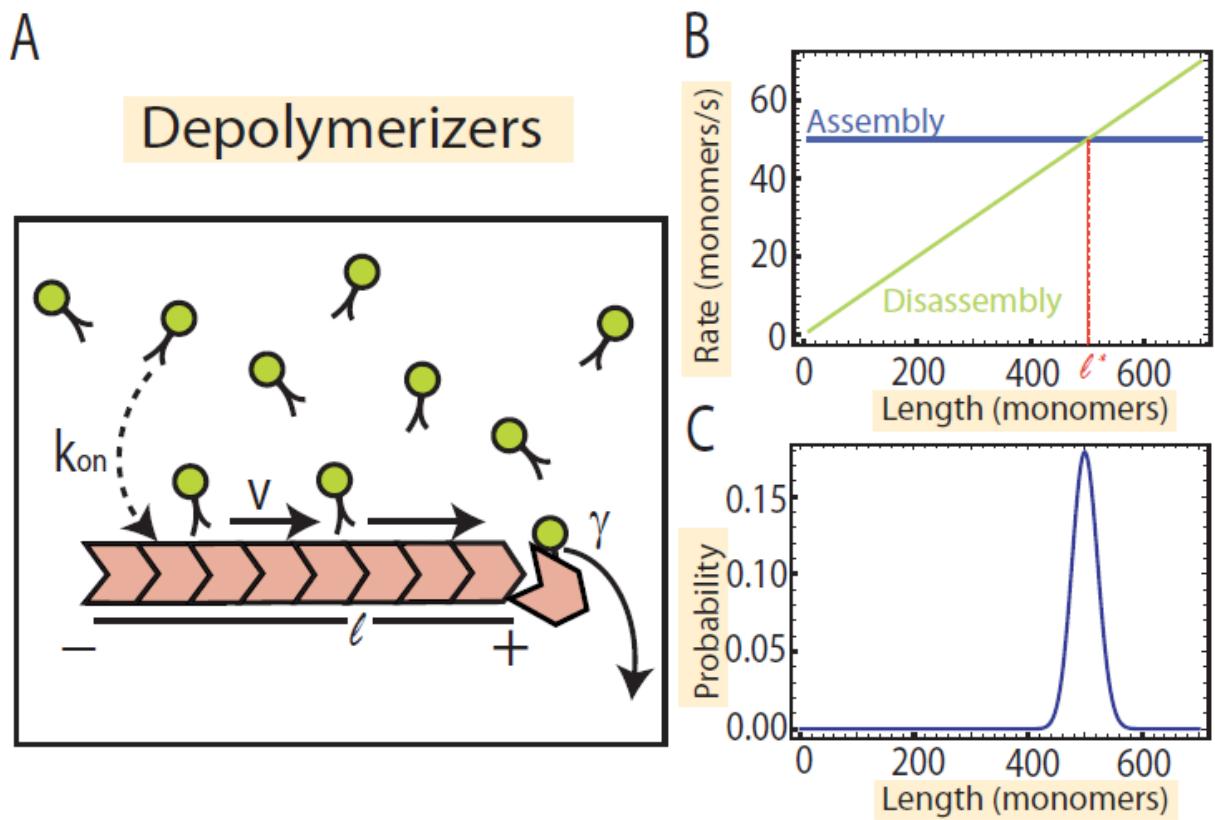

**Figure 9: Active transport of depolymerizers.** (A) A microtubule acts as a landing pad for the motor proteins (green), which reach it at a rate $k_{on}$. Once the motors are on the microtubule, they walk towards the plus end of the microtubule with a rate $v$. When they reach the end, they fall off from the plus end and take a tubulin dimer (red) with them. (B) The disassembly rate is length dependent ($\gamma = \gamma' l$) whereas the growth rate is constant. The two rate curves intersect at a steady-state length given by $l^*$. (C) Steady state length distribution in the regime $r > \gamma'$, i.e., when the rate of growth is greater than the disassembly rate, is peaked. For illustration purposes we used $\frac{r}{\gamma'} = 500$ monomers. steady state mean length.

*2.3.1.2 Length control of microtubules*

A recent study used single molecule analysis to study the effects of the kinesin motor Kip3 on microtubule dynamics (49). It was observed that Kip3 disassembles microtubules exclusively at the plus end and it does so in a length-dependent manner; longer microtubules get shortened faster than shorter ones. These conclusions were reached by doing experiments in vitro with fluorescently labeled microtubules under a TIRF microscope.

The study reported rates for assembly and disassembly as $r$=1µm/min, and $\gamma'$=0.075 /min (at [Kip3]= 3.3 nM) (49). The expected steady state length is therefore $r/\gamma' = 13$ µm (**Figure S1 A**). An increase in the [Kip3] corresponds to an increase in the disassembly rate of the microtubule, and thus from our analysis we expect the mean of the length distribution to shift towards lower microtubule length (**Figure S1 B** ), which is in qualitative agreement with experimental observations.

*2.3.1.3 Key assumptions and additional comments*

For the calculation in section 2.3.1.1, we assumed that the concentration of motors is small enough, so the capture rate of motors on the microtubule is $k_{on}^0$ and depends only on the concentration of motors i.e. $k_{on}^0 = k_{on}'\,[motors]$. But in reality this rate will depend on the number of binding sites on the microtubule. (See SI 5.1 and 5.4 for detailed calculations).

However our simplified picture is valid as long as $k_{on}^0 \ll v\, N_{max}$, where $v$ is the rate at which the motors walk on the microtubule and $N_{max}$ is the maximum number of binding sites. This condition is satisfied, for example, when the relative occupancy of the motors on the microtubule lattice is small.

A few groups have explored this problem in more detail by taking into account the traffic jams that can develop at large numbers of depolymerizing motor proteins walking along the microtubule, and their findings do not qualitatively alter the simple picture presented here (26, 36, 44).

Another phenomenon that we do not consider in the simple model is spontaneous disassembly of microtubules. This seems to be a reasonable assumption for this particular *in vitro* experiment where the rate of Kip3 movement on microtubule is around 4 μm/min. This rate is much faster than the spontaneous rate of microtubule disassembly which is 0.03 μm/min. So the disassembly of microtubule is mostly Kip3 dependent, assuming that the motor protein is highly processive.

For the described mechanism to work, the rate of Kip3 movement on microtubule needs to be larger than the rate at which the filament grows. If that is not the case, then the motor will never be able to make it to the end quickly enough, and cannot play a meaningful role in disassembly of the microtubule end. In such a case, the disassembly rate will still be length independent and will not be able to control the length of the microtubule.

We are also assuming that all the motors make it to the end of the microtubule, and that all of them disassemble the microtubule end. The effect of motor processivity can be accounted for by including a rate for the dissociation of motors from the microtubule, in which case the depolymerizing activity is inversely proportional to the dissociation rate of the motors.

Here we do not consider the dynamic instability of microtubules, choosing to focus on the effect of motor proteins on the disassembly of the microtubule instead. Dynamic instability describes the behavior of purified microtubules *in vitro*, and some microtubules *in vivo*, in which they switch stochastically between phases of slow growth and rapid shrinking. It is to be noted that this process by itself cannot produce a microtubule with a peaked length distribution, and instead leads to an exponential distribution of lengths, or unbounded growth (13). One way of introducing length control in the presence of dynamic instability is to have the frequency of catastrophes (the frequency of transitions from microtubule assembly to rapid disassembly) dependent on the length of microtubules, such that longer microtubules would be more susceptible to catastrophe than shorter ones (49). As we discussed earlier in this section, according to the active transport mechanism longer microtubules will have a larger flux depolymerizers to their ends compared to shorter microtubules. Interestingly, it has been speculated that Kip3 motors may increase the catastrophe frequency by destabilizing the GTP cap at the growing microtubule end (15). In fact, higher catastrophe frequencies have been observed for longer microtubules in *Xenopus* egg extracts (10), suggesting the possible action of an actively transported depolymerizer like Kip3. More recently, a study explored how depolymerizers affect microtubule length distributions in the face of dynamic instability (29) and reported a decrease in mean microtubule length and the narrowing of the microtubule length distribution due to a motor-protein dependent increase in the catastrophe frequency.

*2.3.2 Severing*

Another way of modulating the rate of removal of subunits from filament ends is by filament severing. An example of a severing protein is cofilin that binds to the sides of actin filaments and induces breaks by altering filament conformation (**Figure 10A** and **10B**) (1, 41, 45).

We consider the simplest case of a uniform rate of severing along a filament, in other words severing takes place anywhere on the one-dimensional lattice with equal probability. As we can see in **Figure 10A**, the filament consisting of $l$ subunits can be broken into two smaller filaments at any of the $(l - 1)$ positions with an equal rate $s$, for any choice of severing location; the total rate of severing at any location is then $s(l - 1)$ In addition to the severing rate there is $r$, the rate at which subunits are added to a filament. These two rates define the severing mechanism of length control.

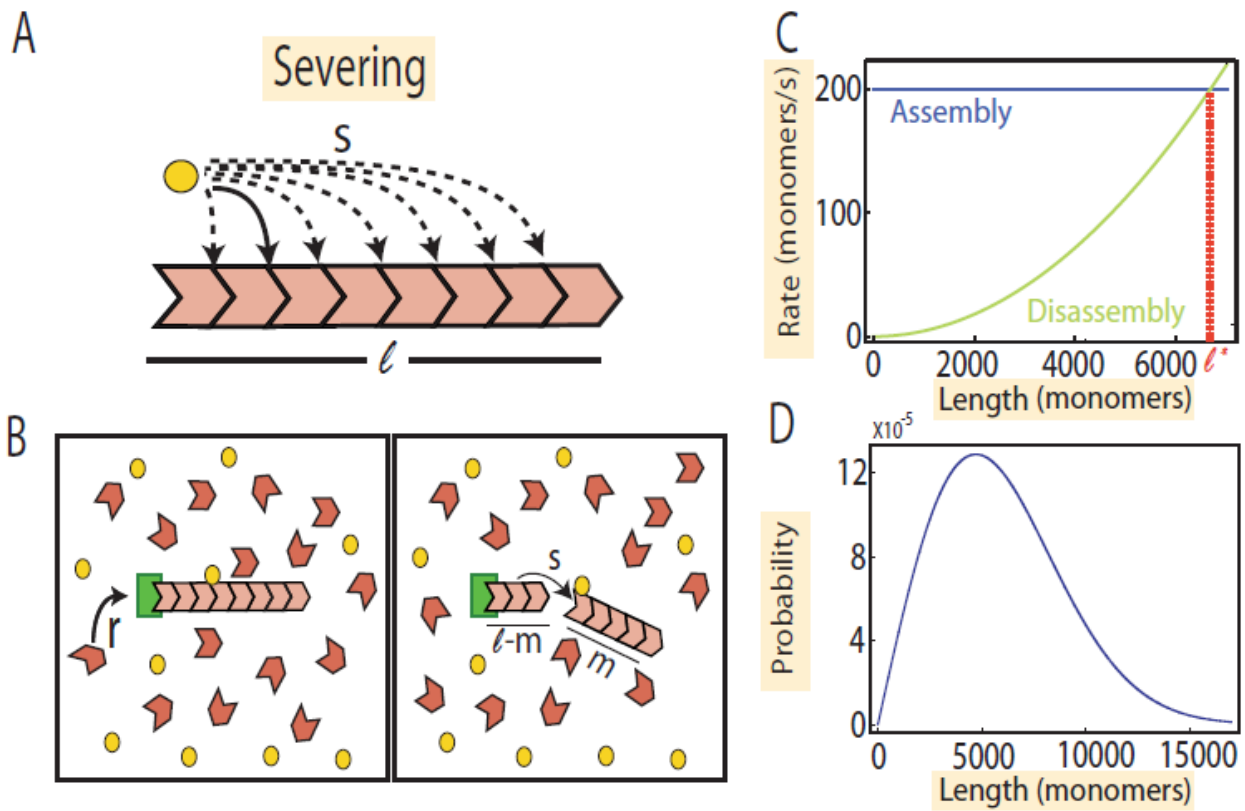

**Figure 10: Severing mechanism of length control.** (A) A filament made up of $l$ subunits has $l - 1$ points of contact between them. Each contact can be severed by a severing protein (yellow) with rate $s$. (B) Monomers are added to a filament by an elongator (green) at a rate $r$. When a filament is severed, a part of the filament is lost resulting in a new filament of length $l - m$

where $m$ is the site of severing. (C) The disassembly rate has a quadratic dependence on the filament length. This coupled with a constant rate of growth can lead to a peaked distribution of filament lengths in steady state. (D) Distribution of filament lengths for $r = 185$ monomer/s (52) and a severing rate $s = 8 \times 10^{-6}$ 1/(monomer s) (1).

When a filament is severed, on average roughly half of the subunits are lost, i.e., they are no longer part of the filament. Note that we only care about the part of the filament which is still connected to the nucleating/assembling center, as in experiments the fragment severed from the filament will rapidly dissolve. Therefore, assuming that severing can occur at any position along the filament that severing protein binds to, the disassembly rate (i.e., rate of subunit loss) is length dependent and scales quadratically with length, $\gamma(l) \sim s\,(l-1)\frac{l}{2} \sim \frac{sl^2}{2}$ (**Figure 10C, Section S1 6.2**).

In actin filaments, the nucleotide state of the filament dictates whether it can be severed or not; newly polymerized actin (ATP or ADP+Pi-actin) in the filament is not a target for severing, whereas aged actin (ADP-actin) is. Because many of the proteins that decorate actin filaments *in vivo* are likely to alter the rate of Pi release, it is not yet clear what fraction of the actin structures in cells is comprised of ADP-actin versus ADP+Pi-actin. For the purposes of our calculations, we have assumed that most of the filament is made up of ADP-actin, and hence is available for severing. As a result, the filaments in our models can effectively be severed at any site.

*2.3.2.1 Master equations for the severing mechanism*

The master equation describing the evolution of $P(l, t)$ in case of severing is

$$\frac{dP(l,t)}{dt} = rP(l-1) - rP(l) + s \sum_{i=l+1}^{\infty} P(i) - s(l-1)P(l). \quad (25)$$

This equation is different from those described so far. In order to produce a filament of length $l$, the filament can either grow from length $l-1$ by adding a subunit with a rate $r$, or it can shrink from any filament having a length larger than $l$ by getting severed with a rate $s$. These terms add to the probability $P(l,t)$. Alternatively, a filament of length $l$ can either get severed in $l-1$ ways with a rate $s$ or add a subunit and become of length $l+1$ with a length independent rate $r$. These terms reduce the probability $P(l,t)$.

At steady state, the probability does not change with time i.e. $\frac{dP(l,t)}{dt} = 0$. Hence from Equation 25, we get $P(l)(r + (l-1)s) = s \sum_{i=l+1}^{\infty} P(i) + P(l-1)r$. Using the normalization condition for the probability $\sum_{i=1}^{\infty} P(i) = \sum_{i=1}^{l} P(i) + \sum_{i=l+1}^{\infty} P(i) = 1$, we obtain

$$P(l)(r + (l-1)s) = s\left(1 - \sum_{i=1}^{l} P(i)\right) + P(l-1)r. \quad (26)$$

Note that in this calculation, we fixed $P(0) = 0$ at steady state, as a condition which is reasonable in the regime where $r > s$. Adding the term $sP(l)$ on both sides of the Equation 26, we can re write this as

$$P(l)(r + ls) = s\left(1 - \sum_{i=1}^{l-1} P(i)\right) + P(l-1)r. \quad (27)$$

Now we solve Equation 27 using the recursion scheme described in Section 1.4 to obtain a closed form solution for $P(l)$, namely

$$P(l) = \frac{l\rho^{l-1}}{(\rho+1)(\rho+2)..(\rho+l)}, \quad (28)$$

where $\rho = \frac{r}{s}$; see SI 6.1 for detailed calculations.

In the regime $\rho < 1$, where the severing rate is larger than the addition rate of subunits, this distribution is a decaying function of the length. The filament gets severed and broken down into fragments faster than subunits are added, hence intuitively it makes sense to have an monotonically decaying function as the steady-state distribution for such a process.

Only when $\rho > 1$, namely, when the addition rate is higher than the rate at which the filament gets severed, is it possible for the filament to have a well defined length with a distribution of lengths that is peaked around the mean (**Figure 10 D**).

From the equation for the mean of a distribution, $\langle l \rangle = \sum_{l=1}^{\infty} l\, P(l)$. we obtain

$$\langle l \rangle = \frac{\rho^{-\rho}\left(\rho^\rho(1+2\rho)+e^\rho(\Gamma[2+\rho]-\Gamma[2+\rho,\rho])\right)}{1+\rho}, \quad (29)$$

where $\Gamma$ is the Gamma function and $\rho = \frac{r}{s}$ (see SI 6.2 for details).

*2.3.2.2 Actin cables in budding yeast*

A actin cable in budding yeast usually grows to a length that approximately equals the diameter of the yeast mother cell, or 5 $\mu$m. The rate of cable assembly is in the range of 0.5-1 $\mu$m/s (52) and the severing rate for actin filaments *in vitro* measured for cofilin is about $10^{-3}$ $\mu m^{-1}$ $s^{-1}$ (1). There are about 3-4 filaments in a cable and we estimate the cofilin-mediated severing rate for the cable to be about 10 times smaller, i.e. $10^{-4}$ $\mu m^{-1}$ $s^{-1}$. Note that there are no measurements of the severing rate *in vivo*, where other factors like Coronin, Aip1, and Srv2/CAP can strongly enhance severing (6, 25). In fact, collectively these co-factors enhance the severing rate 10-30 fold *in vitro*.

Taking our estimated value for the assembly rate, $r = 0.5$ μm/s, and a severing rate $s = 10^{-4}$ μm$^{-1}$ s$^{-1}$, we find a peaked distribution of mean length roughly 90 microns using the exact solution in Equation 43.

Interestingly, this length is more than 15 times the length observed *in vivo*. One would have to change the severing rate by a factor of 400 for agreement with *in vivo* data. Therefore this estimate suggests that severing cannot be the only mechanism of length control in this case, and perhaps an alternate mechanism is working either in conjunction with, or separately from severing to control the length of actin cables. In fact, we recently showed that dampers like Smy1 play an important role in controlling length (39).

*2.3.2.3 Key assumptions and additional comments*

Severing is not limited to actin filaments; proteins like katanin, spastin and fidgetin are known to sever microtubules. However, the severing activity may not be uniform along the filament lattice. A recent paper showed that katanin preferentially targets microtubules lattice defects (9). Even in the case of cofilin, direct visualization of cofilin, actin, and subsequent filament severing events demonstrate that the severing probability is higher at boundaries between bare and cofilin-decorated segments of the filament(11).

For the analysis in Sections 2.3.2.1 and 2.3.2.2, we have assumed that every site on the filament has the same probability of being severed. This in turn depends on what fraction of the filament is composed of ADP+Pi- actin and ADP-actin subunits. It is important to note that severing can control the length of a filament only if a significant fraction of the filament is composed of ADP-actin subunits. If that is not the case and only the ends are ADP-actin, then this will be equivalent to end-depolymerization, discussed earlier in the case of an unregulated filament in Section 2.1. This scenario by itself does not produce a length-dependent disassembly rate. Hence any cellular mechanisms that control the rate of phosphate release on the actin filaments could have a profound impact on severing as a length control mechanism.

Cofilin is not the only protein involved in actin filament severing. In yeast cells, other proteins like Coronin, Srv2/CAP and Aip1 work together with cofilin to sever actin filaments (3, 14, 30). The severing rate increases when all these proteins work together (25). Also, In this analysis, we have not explicitly included the effect of Tropomyosin proteins, which are thought to coat the cables and protect them at least temporarily from cofilin-mediated severing (31, 47). It is still not well understood what the level of actin cable coating by profilin in vivo is. High level of Tropomyosin coating on the cables could make most of the filament inaccessible for severing, possibly resulting in the pruning of the ends of filaments. Again, this type of pruning by severing proteins does not make the disassembly rate length dependent and will not lead to length regulation by severing.

## 3. *Discussion*

The cytoskeleton consists of a wide variety of filamentous structures with characteristic sizes and shapes. In this review, we have described several mechanisms by which cells can control the sizes of these filamentous networks. We introduced a simplified model of a cytoskelton filament which ignores the distinct chemistries of actin and microtubules, yet provides an intuitive view of filament length control. We have resorted to such coarse grained model to emphasize the design principle common to all length control mechanisms described here, namely, the fact that molecular-scale interactions between filaments and associated regulatory proteins leads to a *length dependent rate of filament assembly or disassembly*. For a stable and well defined length to emerge either the assembly rate should diminish with the filament's length or the disassembly rate should increase as the length of the filament increases.

First we described how a finite pool of subunits, actively transported dampers of polymerization, and actively transported monomers to the site of polymerization can produce length-control mechanisms where the assembly rate of the filament slows down as the length of the filament increases. Then we considered the action of actively transported depolymerizing proteins and freely diffusing severing proteins, which control filament length by increasing the rate of disassembly as the length of the filament increases.

For each of these mechanisms we computed the resulting steady-state length distribution. In vivo, such mechanisms almost certainly work in conjunction with each other. For example, in yeast cells, when we delete the actively transported damper protein Smy1, actin cables grow longer but they clearly cannot grow indefinitely, since there is a finite monomer pool. It is still not clear whether the finite monomer pool determines cable length in the absence of Smy1 or some other mechanism limits the growth of cables. Measuring the distributions of filament lengths can then tell us which mechanisms might be at play, and which are not.

*3.1 Experimental signatures of different length control mechanisms*

One of the powerful outcomes of a theoretical analysis like the one presented here is that it can be used to discern between different mechanisms at work. We have argued above that filament length distributions are a signature that distinguishes between different types of length control mechanisms. Earlier we introduced Table 1, where we elucidated all of the different types of mechanisms considered in this review. We listed the resulting closed-form steady-state length distributions for each mechanism, and the corresponding mean length. The change in the moments (e.g., mean and variance) of a distribution as a function of the various parameters that can be experimentally tuned is a specific prediction for each mechanism. This suggests experiments that are designed so that the model parameters can be tuned.

Predictions for all mechanisms considered in this review are summarized in **Figure 11**. A bare filament with length-independent rates of addition and removal of monomers can never lead to a peaked distribution of lengths. If the rate of removal is higher than the rate of monomer addition, it leads to an exponential distribution of lengths. In contrast, a peaked distribution of lengths can be achieved if any of the rates are length-dependent. One way of making rates length-dependent is by having a finite pool of monomers. We obtain a peaked distribution of lengths in this case, and the change of the mean and variance of the distribution with respect to a particular parameter, for example, the total number of monomers, provide experimental predictions for this mechanism (**Figure 11A**).

In the case of a diffusing damper, we can only obtain an exponentially decaying distribution if the resulting mean assembly rate $\bar{r}$ is less than the disassembly rate of the filament. We chose the concentration of dampers ($k_{on}$ in Table 1B) as the parameter to vary, although the binding affinity of the damper for the elongator ($k_{off}$ in Table 1B) could be tuned as well with specific mutations in the binding domain of the damper. The mean and variance of the length distribution decay with increasing damper concentration (**Figure 11B**). Interestingly if the resulting mean assembly rate $\bar{r}$ is larger than the disassembly rate of the filament, then a peaked distribution of lengths can still be obtained if there is a finite pool of monomers (Section 2.2.1).

In the case of length control by an actively transported damper, the assembly rate is length-dependent and results in a filament length distribution given by the expressions in Table 1C in the limit where the switching rates are faster than the rates of assembly and disassembly. In this case, increasing the damper-elongator binding affinity (inversely proportional to $k_{off}$ in Table 1C) increases the mean length, and the variance of the distribution (**Figure 11C**). Alternatively, one can vary the concentration of dampers as the experimental knob; $w$ is then the theoretical parameter that varies in proportion to the concentration of dampers. The length of a filament can also be controlled by having a fixed number of transporters, like in the case of intraflagellar transport, where monomers are carried to the site of assembly. The steady state distribution obtained from this mechanism is a Poisson distribution with the same mean and variance given by expressions in Table 1D. Increasing the number of transporters will increase the mean and variance identically (**Figure 11D**).

For length control by transported depolymerizers we obtain a Poisson distribution with the same mean length and variance, given by the expression in Table 1E. Thus, varying the depolymerizer concentration ($\gamma$ is the corresponding parameter in the model) changes the mean and variance identically (**Figure 11E**).

The process of severing yields a distribution peaked at a mean length given by the expression in Table 1F. Increasing the concentration of severing proteins would increase the severing rate $s$ and hence decrease the mean and variance of the distribution (Figure 11F).

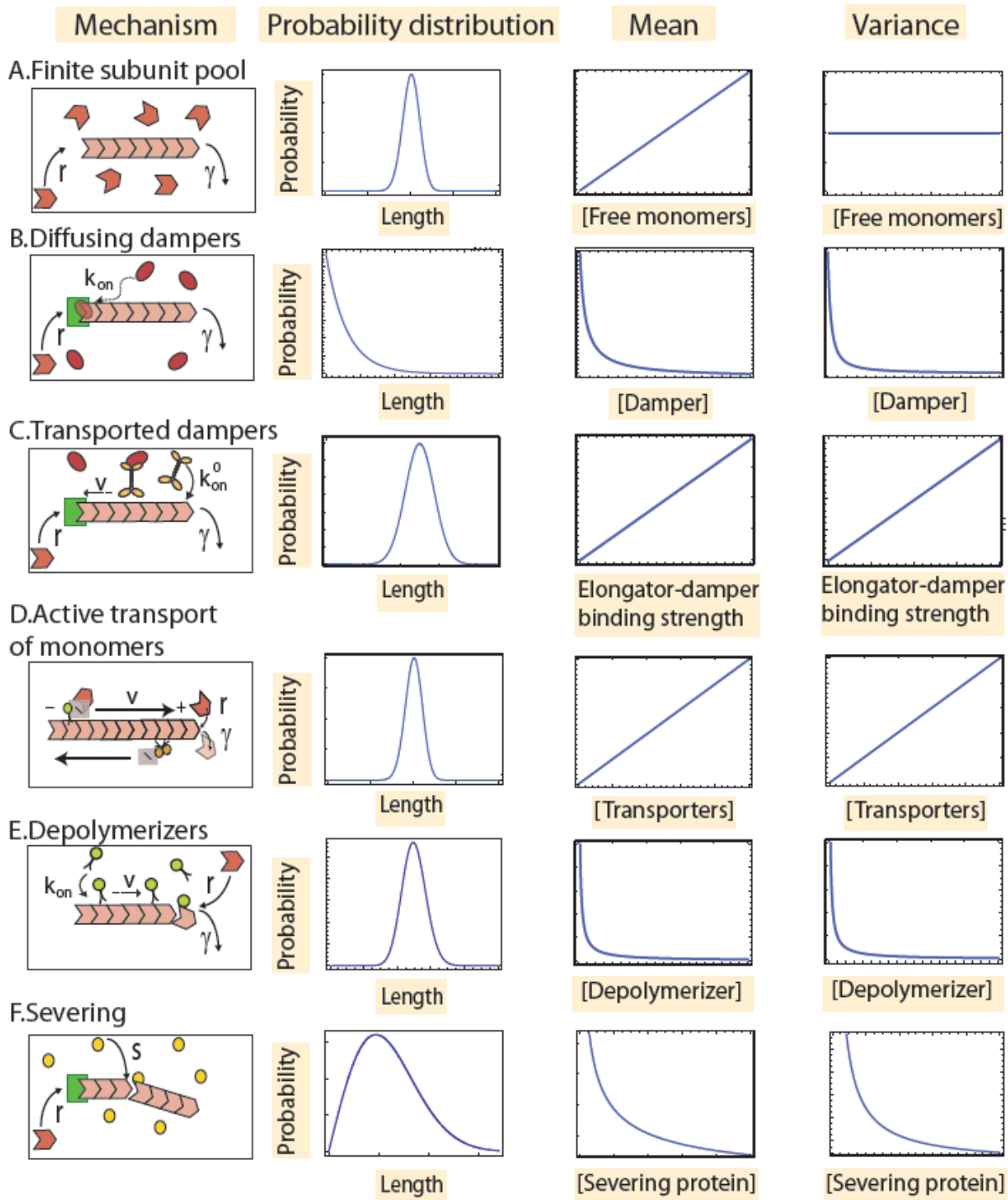

**Figure 11: Experimentally testable signatures of length control mechanisms.** (A) Finite subunit pool. Increasing the available subunit pool ($N_t$ in the analytical expression in Table 1A)

increases the mean length linearly, but the width of distribution is constant. (B) The diffusing damper scenario does not result in length control but instead produces an exponentially decaying distribution of lengths if the resulting mean assembly rate $\bar{r}$ is less than the disassembly rate of the filament. The mean and variance of this distribution will decay with increasing damper concentration ($k_{on}$ is proportional to damper concentration). (C) Length control by an actively transported damper leads to a length distribution peaked at a mean length given by the expression in Table 1C. Increasing the damper-elongator binding strength ($k_{off}$ is proportional to dissociation constant for binding) increases the mean length and the variance of the distribution in the limit of fast switching rates. (D) Length control by the active transport of monomers to the binding site leads to a Poisson distribution, with a mean given by the expression in Table 1D. Increasing the number of transporters will increase the mean and variance of the length distribution. (E) Length control by actively transported depolymerizers yields a Poisson distribution with a mean length given by the expression in Table 1E. Thus, varying the depolymerizer concentration ($\gamma$ is the corresponding model parameter) changes the mean and variance identically. (F) Length control by severing yields a distribution peaked at a mean length given by the expression in Table 1F. Increasing the concentration of severing proteins will increase the severing rate $s$ and decrease the mean and variance of the distribution.

*3.2 The problem of multiple cytoskeleton structures*

A key question in biology is how different-sized cytoskeleton structures coexist in the same cytoplasm while making use of the same building blocks. For example, in yeast cells, actin subunits are used as the common building block to make three different structures: actin cables, patches, and cytokinetic rings. Since the finite pool mechanism is incapable of regulating simultaneously the size of multiple structures (Mohapatra et al, In preparation) other mechanisms have to be employed.

Adding to the complexity of the engineering problem that cells have to deal with, the number of actin subunits distributed between cables and other actin structures can change during the cell cycle. For example, during mitosis, a substantial portion of the actin subunits is abruptly used to construct the cytokinetic ring, depleting it from the available pool for formation of patches and cables. It was recently shown in multiple studies that the distribution of actin subunits between distinct actin structures within a shared cytoplasm can by regulating by actin-binding proteins such as profilin, which favor incorporation of actin into one structure over another (23, 46, 48).

Yeast cells have distinct actin nucleating factors, which establish structures of different geometries. Formins generate structures comprised of linear actin filaments, whereas the Arp2/3 complex forms networks comprised of branched actin filaments. These nucleating factors have to compete for a common pool of monomers, and formins have a disadvantage because of their low cellular concentration. A recent study showed that cells can use small actin binding proteins like profilin and capping proteins to successfully compete with Arp2/3 complex for actin monomers (4, 46, 48). This raises the question, do different cells alter their profilin and capping protein levels under different conditions and/or stages of the cell cycle so as to direct actin monomers to assemble into a particular structure?

*3.3 Size control of non-cytoskeletal structures in the cell*

In this review we have focused on length control of cytoskeletal structures in eukaryotic cells; however, cells also contain numerous non-cytoskeletal structures whose number and size appear to be under tight control(17, 21, 43, 51) . The mechanisms regulating the sizes of these organelles remain largely unknown, but we think that coarse-grained models similar to those described herein could be used to provide new insights into these problems and guide future experiments.

One of the earliest observations of size control was the scaling of the nucleus with the size of red blood cells described by Gulliver in 1875 (17). Recent experiments show that when a nucleus is transferred from a small donor cell (HeLa) into the cytoplasm of larger host cell (Xenopus oocytes), the transferred nucleus expands, responding to the size of the host (22). This observation seems in line with a currently favored view that size control is exerted by a limiting pool of cytoplasmic components (17). Another example is the size control of centrosomes, which organize microtubules during spindle assembly(21). It has been hypothesized that centrosome size is controlled by the recruitment of a limited number of building blocks available for centrosome formation (17, 27). Other organelles besides the nucleus whose size scales according to cell size are mitochondria, which can grow and divide from a finite number of mitochondrial particles, but in addition can undergo fission and fusion. An active feedback and sensing mechanism was proposed to explain the scaling of the size of the mitochondrial network with bud cell size in yeast cells (43), where it was suggested that the cells sense the kinetics of mitochondrial accumulation in the bud and respond to ensure proper levels of mitochondrial accumulation before cell division.

What is remarkable is that even though the above-mentioned structures have very distinct functions, the mechanisms controlling their formation and maintenance seem to be the same and rely on a finite pool of building blocks. Therefore, the principles described in Section 2.2.1 can be used broadly to understand the growth of these structures, and predict their scaling with the size of the compartment in which they are located, and the number of diffusing components(17).

However, we suspect that the finite monomer pool is not the only mechanism at play controlling the size of these structures, and that yet to be discovered proteins regulating the rates of assembly and disassembly might be playing a role as well. One of the key lessons learned above is that size control is achieved when the molecular-scale interactions between regulatory proteins and filaments lead to size dependent assembly or disassembly rate. We speculate that similar principles might be involved in regulating the size of other sub-cellular structures.

The purpose of this review has been to place length-control mechanisms described thus far on a common footing using simple, coarse-grained models, so that their predictions can be compared and contrasted. Our goal is to inspire a new wave of experiments that directly measure length distributions of cytoskeleton filaments so as to discern between different mechanisms being used by cells to control the length of these filaments. Furthermore, we believe that these same principles may extend to other, non-cytoskeletal structures for example, centrosomes, nucleoli particles and mitochondria, whose size seems to be controlled by cells in response to different intracellular (cell cycle dependent) and extracellular (growth condition dependent) cues.

*Acknowledgments*

We wish to thank the Brandeis Cable Club (Julian Eskin, Sal Alioto, Brian Graziano, Mikael Garabedian), for many stimulating discussions about cytoskeletal length regulation and the biological mechanisms underlying them. Discussions with Wallace Marshall, Stephanie Weber, David Kovar, Fred Chang, and Arjun Raj about general issues regarding organelle size control were also very helpful.

This work was supported by the NSF through grant DMR-1206146 (J.K.) and MRSEC-1420382 (B.G. and J.K.), and by the NIH DP1OD000217 (R.P.), and R01GM085286 (R.P.),

R01GM083137 (B.G.). We are grateful to the Burroughs-Wellcome Fund for its support of the Physiology Course at the Marine Biological Laboratory, where part of the work on this review was done, and for a post-course research grant (L.M.).

*Literature Cited*

1. Andrianantoandro E, Pollard TD. 2006. Mechanism of actin filament turnover by severing and nucleation at different concentrations of adf/cofilin. *Mol. Cell*. 24(1):13–23

2. Avasthi P, Marshall WF. 2012. Stages of ciliogenesis and regulation of ciliary length. *Differ. Res. Biol. Divers.* 83(2):S30–42

3. Balcer HI, Goodman AL, Rodal AA, Smith E, Kugler J, et al. 2003. Coordinated regulation of actin filament turnover by a high-molecular-weight srv2/cap complex, cofilin, profilin, and aip1. *Curr. Biol. CB*. 13(24):2159–69

4. Burke TA, Christensen JR, Barone E, Suarez C, Sirotkin V, Kovar DR. 2014. Homeostatic actin cytoskeleton networks are regulated by assembly factor competition for monomers. *Curr. Biol. CB*. 24(5):579–85

5. Buttery SM, Yoshida S, Pellman D. 2007. Yeast formins bni1 and bnr1 utilize different modes of cortical interaction during the assembly of actin cables. *Mol. Biol. Cell*. 18(5):1826–38

6. Chaudhry F, Breitsprecher D, Little K, Sharov G, Sokolova O, Goode BL. 2013. Srv2/cyclase-associated protein forms hexameric shurikens that directly catalyze actin filament severing by cofilin. *Mol. Biol. Cell*. 24(1):31–41

7. Chesarone-Cataldo M, Guérin C, Yu JH, Wedlich-Soldner R, Blanchoin L, Goode BL. 2011. The myosin passenger protein smy1 controls actin cable structure and dynamics by acting as a formin damper. *Dev. Cell*. 21(2):217–30

8. Chesarone M, Gould CJ, Moseley JB, Goode BL. 2009. Displacement of formins from growing barbed ends by bud14 is critical for actin cable architecture and function. *Dev. Cell*. 16(2):292–302

9. Díaz-Valencia JD, Morelli MM, Bailey M, Zhang D, Sharp DJ, Ross JL. 2011. Drosophila katanin-60 depolymerizes and severs at microtubule defects. *Biophys. J.* 100(10):2440–49

10. Dogterom M, Félix MA, Guet CC, Leibler S. 1996. Influence of m-phase chromatin on the anisotropy of microtubule asters. *J. Cell Biol.* 133(1):125–40

11. Elam WA, Kang H, De la Cruz EM. 2013. Biophysics of actin filament severing by cofilin. *FEBS Lett.* 587(8):1215–19

12. Erlenkämper C, Kruse K. 2009. Uncorrelated changes of subunit stability can generate length-dependent disassembly of treadmilling filaments. *Phys. Biol.* 6(4):046016

13. Fygenson DK, Braun E, Libchaber A. 1994. Phase diagram of microtubules. *Phys. Rev. E*. 50(2):1579–88

14. Gandhi M, Achard V, Blanchoin L, Goode BL. 2009. Coronin switches roles in actin disassembly depending on the nucleotide state of actin. *Mol. Cell*. 34(3):364–74

15. Gardner MK, Zanic M, Howard J. 2013. Microtubule catastrophe and rescue. *Curr. Opin. Cell Biol.* 25(1):14–22

16. Ghaemmaghami S, Huh W-K, Bower K, Howson RW, Belle A, et al. 2003. Global analysis of protein expression in yeast. *Nature*. 425(6959):737–41

17. Goehring NW, Hyman AA. 2012. Organelle growth control through limiting pools of cytoplasmic components. *Curr. Biol. CB*. 22(9):R330–39

18. Goshima G, Wollman R, Stuurman N, Scholey JM, Vale RD. 2005. Length control of the metaphase spindle. *Curr. Biol. CB*. 15(22):1979–88

19. Gould CJ, Chesarone-Cataldo M, Alioto SL, Salin B, Sagot I, Goode BL. 2014. Saccharomyces cerevisiae kelch proteins and bud14 protein form a stable 520-kda formin regulatory complex that controls actin cable assembly and cell morphogenesis. *J. Biol. Chem.* 289(26):18290–301

20. Graziano BR, Yu H-YE, Alioto SL, Eskin JA, Ydenberg CA, et al. 2014. The f-bar protein hof1 tunes formin activity to sculpt actin cables during polarized growth. *Mol. Biol. Cell*. 25(11):1730–43

21. Greenan G, Brangwynne CP, Jaensch S, Gharakhani J, Jülicher F, Hyman AA. 2010. Centrosome size sets mitotic spindle length in caenorhabditis elegans embryos. *Curr. Biol. CB*. 20(4):353–58

22. Gurdon JB. 1976. Injected nuclei in frog oocytes: fate, enlargement, and chromatin dispersal. *J. Embryol. Exp. Morphol.* 36(3):523–40

23. Henty-Ridilla JL, Goode BL. 2015. Global resource distribution: allocation of actin building blocks by profilin. *Dev. Cell*. 32(1):5–6

24. Ishikawa H, Marshall WF. 2011. Ciliogenesis: building the cell's antenna. *Nat. Rev. Mol. Cell Biol.* 12(4):222–34

25. Jansen S, Collins A, Chin SM, Ydenberg CA, Gelles J, Goode BL. 2015. Single-molecule imaging of a three-component ordered actin disassembly mechanism. *Nat. Commun.* 6:7202

26. Johann D, Erlenkämper C, Kruse K. 2012. Length regulation of active biopolymers by molecular motors. *Phys. Rev. Lett.* 108(25):258103

27. Kirkham M, Müller-Reichert T, Oegema K, Grill S, Hyman AA. 2003. Sas-4 is a c. elegans centriolar protein that controls centrosome size. *Cell*. 112(4):575–87

28. Kovar DR, Pollard TD. 2004. Insertional assembly of actin filament barbed ends in association with formins produces piconewton forces. *Proc. Natl. Acad. Sci. U. S. A.* 101(41):14725–30

29. Kuan H-S, Betterton MD. 2013. Biophysics of filament length regulation by molecular motors. *Phys. Biol.* 10(3):036004

30. Kueh HY, Charras GT, Mitchison TJ, Brieher WM. 2008. Actin disassembly by cofilin, coronin, and aip1 occurs in bursts and is inhibited by barbed-end cappers. *J. Cell Biol.* 182(2):341–53

31. Kuhn TB, Bamburg JR. 2008. Tropomyosin and adf/cofilin as collaborators and competitors. *Adv. Exp. Med. Biol.* 644:232–49

32. Lin HW, Schneider ME, Kachar B. 2005. When size matters: the dynamic regulation of stereocilia lengths. *Curr. Opin. Cell Biol.* 17(1):55–61

33. Manor U, Kachar B. 2008. Dynamic length regulation of sensory stereocilia. *Semin. Cell Dev. Biol.* 19(6):502–10

34. Marshall WF, Qin H, Rodrigo Brenni M, Rosenbaum JL. 2005. Flagellar length control system: testing a simple model based on intraflagellar transport and turnover. *Mol. Biol. Cell*. 16(1):270–78

35. Marshall WF, Rosenbaum JL. 2001. Intraflagellar transport balances continuous turnover of outer doublet microtubules: implications for flagellar length control. *J. Cell Biol.* 155(3):405–14

36. Melbinger A, Reese L, Frey E. 2012. Microtubule length regulation by molecular motors. *Phys. Rev. Lett.* 108(25):258104

37. Mitchell DR. 2004. Speculations on the evolution of 9+2 organelles and the role of central pair microtubules. *Biol. Cell Auspices Eur. Cell Biol. Organ.* 96(9):691–96

38. Mitchell DR. 2007. The evolution of eukaryotic cilia and flagella as motile and sensory organelles. *Adv. Exp. Med. Biol.* 607:130–40

39. Mohapatra L, Goode BL, Kondev J. 2015. Antenna mechanism of length control of actin cables. *PLoS Comput. Biol.* 11(6):e1004160

40. Niwa S, Nakajima K, Miki H, Minato Y, Wang D, Hirokawa N. 2012. Kif19a is a microtubule-depolymerizing kinesin for ciliary length control. *Dev. Cell*. 23(6):1167–75

41. Pavlov D, Muhlrad A, Cooper J, Wear M, Reisler E. 2007. Actin filament severing by cofilin. *J. Mol. Biol.* 365(5):1350–58

42. Pollard TD. 1986. Rate constants for the reactions of atp- and adp-actin with the ends of actin filaments. *J. Cell Biol.* 103(6 Pt 2):2747–54

43. Rafelski SM, Viana MP, Zhang Y, Chan Y-HM, Thorn KS, et al. 2012. Mitochondrial network size scaling in budding yeast. *Science*. 338(6108):822–24

44. Reese L, Melbinger A, Frey E. 2014. Molecular mechanisms for microtubule length regulation by kinesin-8 and xmap215 proteins. *Interface Focus*. 4(6):20140031

45. Roland J, Berro J, Michelot A, Blanchoin L, Martiel J-L. 2008. Stochastic severing of actin filaments by actin depolymerizing factor/cofilin controls the emergence of a steady dynamical regime. *Biophys. J.* 94(6):2082–94

46. Rotty JD, Wu C, Haynes EM, Suarez C, Winkelman JD, et al. 2015. Profilin-1 serves as a gatekeeper for actin assembly by arp2/3-dependent and -independent pathways. *Dev. Cell*. 32(1):54–67

47. Skau CT, Kovar DR. 2010. Fimbrin and tropomyosin competition regulates endocytosis and cytokinesis kinetics in fission yeast. *Curr. Biol. CB*. 20(16):1415–22

48. Suarez C, Carroll RT, Burke TA, Christensen JR, Bestul AJ, et al. 2015. Profilin regulates f-actin network homeostasis by favoring formin over arp2/3 complex. *Dev. Cell*. 32(1):43–53

49. Varga V, Helenius J, Tanaka K, Hyman AA, Tanaka TU, Howard J. 2006. Yeast kinesin-8 depolymerizes microtubules in a length-dependent manner. *Nat. Cell Biol.* 8(9):957–62

50. Varga V, Leduc C, Bormuth V, Diez S, Howard J. 2009. Kinesin-8 motors act cooperatively to mediate length-dependent microtubule depolymerization. *Cell*. 138(6):1174–83

51. Weber SC, Brangwynne CP. 2015. Inverse size scaling of the nucleolus by a concentration-dependent phase transition. *Curr. Biol.* 25(5):641–46

52. Yu JH, Crevenna AH, Bettenbühl M, Freisinger T, Wedlich-Söldner R. 2011. Cortical actin dynamics driven by formins and myosin v. *J. Cell Sci.* 124(Pt 9):1533–41

**Supplementary Information for Design Principles of Length Control of Cytoskeletal Structures**

*1. Unregulated filament*

*1.1 Time- dependent mean filament length in the regime $r > \gamma$*

The master equation for an unregulated filament is given by Equation 3 in Section 1.3 in main text, namely

$$\frac{dP(l,t)}{dt} = r\, P(l-1) - rP(l) + \gamma\, P(l+1) - \gamma\, P(l). \quad (1)$$

The equation for the average length is obtained by multiplying both sides of Equation 1 by $l$ and summing over all possible lengths, namely,

$$\frac{d\langle l \rangle}{dt} = \sum_{l=0}^{\infty} \frac{d}{dt} l\, P(l,t)$$

$$= r \sum_{l=0}^{\infty} l\, P(l-1) - r \sum_{l=0}^{\infty} l\, P(l) + \gamma \sum_{l=0}^{\infty} l\, P(l+1) - \gamma \sum_{l=0}^{\infty} l\, P(l). \quad (2)$$

The terms on the right hand side of this equation can all be rewritten as moments of the $P(l,t)$ disribution. Using the definition of moments, $\langle l^n \rangle = \sum_{l=0}^{\infty} l^n\, P(l,t)$, the second and fourth term are simply $-r\langle l \rangle$ and $-\gamma\langle l \rangle$ respectively. The first and third terms require a bit more attention.

We begin with the first term $r \sum_{l=0}^{\infty} l\, P(l-1)$ on the RHS of equation 2. Since $P(-1) = 0$, we can write this terms as $r \sum_{l=1}^{\infty} l\, P(l-1)$. Changing the variables using $m = l - 1$, we obtain

$$r \sum_{l=1}^{\infty} l\, P(l-1) = r \sum_{m=0}^{\infty} (m+1) P(m). \quad (3)$$

Next we consider the third term $\gamma \sum_{l=0}^{\infty} l\, P(l+1)$ in the RHS of equation 2, where, by adding and subtracting $P(0)$, we can write this term as $\gamma \left( \sum_{l=-1}^{\infty} l\, P(l+1) + P(0) \right)$. Changing the variables $m = l + 1$, we obtain

$$\gamma \sum_{l=-1}^{\infty} l\, P(l+1) = \gamma \left( \sum_{m=0}^{\infty} (m-1) P(m) + P(0) \right). \quad (4)$$

Substituting Equation 3 and 4 in Equation 2 and using the definition of moments $\langle l^n \rangle = \sum_{l=0}^{\infty} l^n\, P(l,t)$, we obtain

$$\frac{d}{dt} \langle l(t) \rangle = r - \gamma + \gamma\, P(0). \quad (5)$$

When $r > \gamma$, the filament will grow more than shrink and hence at later times, $P(0) \sim 0$. Hence, Equation 5 can thus be solved to obtain $\langle l(t) \rangle = (r - \gamma) t$, which implies that the filament will grow linearly with time and there is no steady state. Note that here we are considering an infinite pool of monomers. We consider the effect of a finite pool in Section 2.2.1.

*1.2 Solving master equations for $r < \gamma$*

Consider a filament which adds and subtracts subunits at rates $r$ and $\gamma$ respectively. We are interested in computing the steady state probability distribution of lengths i.e $P(l)$, where $l$ is the length of the filament in the regime where $r < \gamma$. We start by writing down the master equation for the states $P(l, t)$, namely

$$\frac{dP(l,t)}{dt} = r\,P(l-1) - rP(l) + \gamma\,P(l+1) - \gamma\,P(l). \qquad (6)$$

The master equation for $P(0, t)$, which is the probability of being at zero length at time $t$,

$$\frac{dP(0,t)}{dt} = \gamma\,P(1) - r\,P(0). \qquad (7)$$

needs to be considered separately since the transition of $P(0)$ decaying into $P(-1)$ is not allowed. At steady state, $\frac{dP(0,t)}{dt} = 0$. We can use the steady-state equation for $P(0)$ to obtain a general expression for steady state distribution $P(l)$ by using recursion, namely

$$P(1) = \frac{r}{\gamma} P(0). \qquad (8)$$

Also, at steady state $\frac{dP(l,t)}{dt} = 0$ and Equation 6 becomes

$$r\,P(l-1) + \gamma\,P(l+1) = (r+\gamma)P(l). \qquad (9)$$

Substituting $l = 1$ in Equation 9, we obtain

$$r\,P(0) + \gamma\,P(2) = (r+\gamma)P(1). \qquad (10)$$

Now substituting Equation 8, we get a simple relationship between $P(2)$ and $P(0)$ namely, $P(2) = \left(\frac{r}{\gamma}\right)^2 P(0)$. By using this recursion scheme of recursion, we obtain

$$P(l) = \left(\frac{r}{\gamma}\right)^l P(0).$$

## 2. Finite monomer pool

### 2.1 Solving the master equation

We want to solve the master equation for an individual filament in a finite pool of subunits, namely

$$\frac{dP(l,t)}{dt} = r(l-1)P(l-1) - r(l)P(l) + \gamma\, P(l+1) - \gamma\, P(l). \quad (11)$$

Again, at steady state, we have $\frac{dP(l,t)}{dt} = 0$. Now let us consider the equation for $P(0)$, the probability of having zero subunits. Since $P(-1) = 0$ and the $P(0)$ state cannot decay, the equation for $P(0)$ is given by

$$\frac{dP(0,t)}{dt} = \gamma\, P(1) - r(0)P(0)\,. \quad (12)$$

In other words, the only allowed transitions for $P(0)$ are the decay from the state P(1) and growth to the state $P(1)$. The decay from the state $P(1)$ where a filament with 1 subunit, loses the subunit at a rate $\gamma$ and has no subunits. This term adds to $P(0,t)$. The growth to the state $P(1)$, where a filament with no subunits, adds a subunit at a rate $r(0)$ and now has 1 subunit. This term reduces $P(0,t)$.

Once again at steady state, $\frac{dP(0,t)}{dt} = 0$. Substituting in Equation 12, we obtain $P(1) = \frac{r(0)}{\gamma} P(0)$ which is $\frac{r' \times (N_f)}{\gamma} P(0)$. For $l = 1$, from Equation 12, $r(0) P(0) + \gamma P(2) = (r(1) + \gamma)P(1)$. We can substitute $P(1)$ in terms of $P(0)$ to obtain an expression for $P(2)$, namely

$$(r' \times N_t)P(0) + \gamma P(2) = (r' \times (N_t - 1)) + \gamma) \frac{r' \times N_t}{\gamma} P(0). \quad (13)$$

Rearranging the terms in equation 3, we can obtain P(2) as a function of P(0) as

$$P(2) = \frac{r' \times N_t}{\gamma} \left( \frac{r' \times (N_t - 1)}{\gamma} \right) P(0). \quad (14)$$

Using $\frac{N_t!}{(N_t-2)!} = N_t \times (N_t - 1)$, we can rewrite

$$P(2) = \left( \frac{r'}{\gamma} \right)^2 \frac{N_t!}{N_t - 2)!} P(0). \quad (15)$$

Repeating this recursively for $l = 3$ and so on, we can get a general expression for $P(l)$ in terms of $P(0)$ as

$$P(l) = \left( \frac{r'}{\gamma} \right)^l \frac{N_t!}{(N_t - l)!} P(0), \quad (16)$$

*2.2 Equation for the mean*

We compute the mean of this distribution by multiplying the master equation (Equation 11) by $l$ and summing over all possible lengths, namely

$$\frac{d\langle l \rangle}{dt} = \sum_{l=0}^{N_t} \frac{d}{dt} l\, P(l,t) = \sum_{l=0}^{N_t} r(l-1)\, l\, P(l-1) - \sum_{l=0}^{N_t} r(l)\, l\, P(l) + \gamma \sum_{l=0}^{N_t} l\, P(l+1) - \gamma \sum_{l=0}^{N_t} l\, P(l). \quad (17)$$

The length of each filament can theoretically vary from 0 to $N_t$, where all the monomers are incorporated in the filament itself. we consider each of the sums on the right hand side separately. We begin with the first term $\sum_{l=0}^{N_t} r(l-1)\, l\, P(l-1)$ on the RHS of equation 17. Since $P(-1) = 0$, we can write this terms as $\sum_{l=1}^{N_t} r(l-1)\, l\, P(l-1)$. Changing the variables using $m = l - 1$, we obtain

$$\sum_{l=1}^{N_t} r(l-1)\, l\, P(l-1) = \sum_{m=0}^{N_t} r(m)(m+1) P(m). \quad (18)$$

Note that in Equation 18, we are assuming that $N_t$ is large enough that $N_t - 1 \sim N_t$. Next we consider the third term $\gamma \sum_{l=0}^{N_t} l\, P(l+1)$ in the RHS of equation 17. By adding and subtracting $P(0)$, we can write this term as $\gamma \left( \sum_{l=-1}^{N_t} l\, P(l+1) + P(0) \right)$. Changing the variables $m = l + 1$, we obtain

$$\gamma \sum_{l=-1}^{N_t} l\, P(l+1) = \gamma \sum_{m=0}^{N_t} (m-1) P(m). \quad (19)$$

Again, note that in Equation 14, we are assuming that $N_t$ is large enough that $N_t + 1 \sim N_t$. $P(0)$ is 1 at early times but at later times, for $r' > \gamma$, $P(0) \sim 0$. Substituting Equation 18 and 19 in Equation 17 and using the definition of moments $\langle l^n \rangle = \sum_{l=0}^{N_t} l^n\, P(l,t)$, we obtain

$$\frac{d}{dt} \langle l \rangle = r'\, N_t \langle l \rangle - r' \langle l^2 \rangle + r'\, N_t - r' \langle l \rangle - r'\, N_t \langle l \rangle + r' \langle l^2 \rangle + \gamma \langle l^2 \rangle - \gamma - \gamma \langle l^2 \rangle, \quad (20)$$

which further simplifies to $\frac{d}{dt}\langle l \rangle = r' N_t - r'\langle l \rangle - \gamma$. At steady state, $\frac{d}{dt}\langle l \rangle = 0$ and we obtain $\langle l \rangle = N_t - \frac{\gamma}{r'}$.

*2.3 Equation for the variance*

For calculating variance, we multiply the master equation in Equation 11 with $l^2$ and sum over all possible lengths, namely

$$\sum_{l=0}^{N_t} \frac{d}{dt} l^2 P(l,t) = \sum_{l=0}^{N_t} r(l-1)l^2 P(l-1) - \sum_{l=0}^{N_t} r(l)l^2 P(l) + \gamma \sum_{l=0}^{N_t} l^2 P(l+1) - \gamma \sum_{l=0}^{N_f} l^2 P(l). \quad (21)$$

Let us consider the first term $\sum_{l=0}^{N_t} r(l-1) l^2 P(l-1)$ in the RHS of equation 21. Since $P(-1) = 0$, we can write this terms as $\sum_{l=1}^{N_t} r(l-1) l^2 P(l-1)$. Changing the variables $m = l - 1$, we obtain

$$\sum_{l=1}^{N_t} r(l-1) l^2 P(l-1) = \sum_{m=0}^{N_t} r(m)(m+1)^2 P(m). \quad (22)$$

Now let us consider the third term $\gamma \sum_{l=0}^{N_t} l^2 P(l+1)$ in the RHS of equation 21. Adding and subtracting $P(0)$, we can write this terms as $\gamma \left( \sum_{l=-1}^{N_t} l^2 P(l+1) - P(0) \right)$. Changing the variables $m = l + 1$, we obtain

$$\gamma \sum_{l=-1}^{N_t} l^2 P(l+1) = \gamma \sum_{m=0}^{N_t} (m-1)^2 P(m). \quad (23)$$

Once again, Note that in Equation 22 and 23, we are assuming that $N_t$ is large enough that $N_t \pm 1 \sim N_t$ and $P(0) \sim 0$. Substituting Equation 22 and 23 in Equation 21, we get

$$\frac{d}{dt}\sum_{l=0}^{N_t} l^2\, P(l,t) = \sum_{m=0}^{N_t} r' \times (N_t - m)(m+1)^2 P(m) - \sum_{l=0}^{N_t} r' \times N_t l^2\, P(l) + \gamma \sum_{m=0}^{N_t} (m-1)^2 P(m) - \gamma \sum_{l=0}^{N_t} l^2\, P(l). \quad (24)$$

Once again using the definition of moments $\langle l^n \rangle = \sum_{l=0}^{N_f} l^n\, P(l,t)$, we obtain

$$\frac{d}{dt}\langle l^2 \rangle = r'\, N_t \langle l^2 \rangle - r'\, \langle l^3 \rangle + r'\, N_t - r'\langle l \rangle + 2\, r' N_t \langle l \rangle - 2\, r' \langle l^2 \rangle - r'\, N_t \langle l^2 \rangle + r'\, \langle l^3 \rangle + \gamma \langle l^2 \rangle$$
$$- \gamma - 2\gamma \langle l \rangle - \gamma \langle l^2 \rangle. \quad (25)$$

which further simplifies to

$$\frac{d}{dt}\langle l^2 \rangle = r'\, N_t - r'\langle l \rangle + 2r' N_t \langle l \rangle - 2r'\, \langle l^2 \rangle + \gamma - 2\gamma \langle l \rangle. \quad (26)$$

At steady state, $\frac{d}{dt}\langle l^2 \rangle = 0$, Hence $2r'\, \langle l^2 \rangle = r'\, N_t - r'\langle l \rangle + 2r' N_t \langle l \rangle + \gamma - 2\gamma \langle l \rangle$. Substituting $\langle l \rangle = N_t - \frac{\gamma}{r'}$, we obtain

$$2r'\, \langle l^2 \rangle = r'\left(\langle l \rangle + \frac{\gamma}{r'}\right) - r'\langle l \rangle + 2r'\left(\langle l \rangle + \frac{\gamma}{r'}\right)\langle l \rangle + \gamma - 2\gamma \langle l \rangle, \quad (27)$$

which further simplifies to $2r'\, \langle l^2 \rangle - 2r'\langle l \rangle^2 = 2\gamma,$ yielding an expression for the variance of the distribution as

$$\langle l^2 \rangle - \langle l \rangle^2 = \frac{\gamma}{r'}. \quad (28)$$

*3. Directed dampers*

*3.1 Solving the master equation*

The average time that the filament spends in the ON state, when the elongator is active and the filament is growing at rate $r$, is $1/k_{on}(l)$, while the average time the filament spends in the OFF state is $1/k_{off}$. Since we assume that the rate of growth in the OFF state is zero, the average rate of polymerization is $\bar{r}(l) = r\left(\frac{k_{off}}{k_{off}+k_{on}(l)}\right)$, where the factor appearing in parenthesis is the fraction of time that the filament spends in the ON state. In the fast switching limit, the filament can be assumed to have an instantaneous polymerization rate, $\bar{r}$, which is length dependent since $k_{on}(l) = wl$, and a depolymerisation rate $\gamma$. Using the detailed balance condition $P(l)\bar{r}(l) = \gamma P(l+1)$ we obtain

$$P(l)\, r\, \frac{k_{off}}{k_{off} + wl} = \gamma\, P(l+1). \qquad (29)$$

Writing equation 29 for $l = 0$, we get an equation for $P(1)$ in terms of $P(0)$, the probability of zero subunits present at the elongator, namely $P(1) = \frac{r}{\gamma} P(0)$. Writing equation 29 for $l = 1$, we get an equation for $P(2)$ in terms of $P(0)$ and repeating the same scheme for $l = 2, 3$ and so on, we can get an expression for $P(l)$ in terms of $P(0)$, namely

$$P(l) = \left(\frac{r}{\gamma}\right)^l \prod_{i=0}^{l-1} \left(\frac{k_{off}}{k_{off} + iw}\right) P(0). \qquad (30)$$

We use the normalization condition for $P(l)$ i.e. $\sum_{l=0}^{\infty} P(l) = 1$ to obtain $P(0)$, which then gives us an analytic formula for the length distribution

$$P(l) = \left(\frac{r}{d}\right)^l \frac{\left(k_{off}/w\right)^{l-1}}{\left(\frac{\Gamma\left(\frac{k_{off}}{w} + l\right)}{\Gamma(l-1)}\right)} \left(\frac{e^{\frac{k_{off}r}{\gamma w}} k_{off}\, r(k_{off} - w) \left(\frac{k_{off}\, r}{\gamma w}\right)^{-\left(\frac{k_{off}}{w}\right)} \left(\Gamma\left[\frac{k_{off} - w}{w}\right] - \Gamma\left[-1 + \frac{k_{off}}{w}, \frac{k_{off}\, r}{\gamma w}\right]\right)}{\gamma w^2}\right)^{-1}, \qquad (31)$$

where $\Gamma(x)$ is the gamma function.

*3.2 Estimation of parameters for actin cables in budding yeast*

This mechanism of active transport of dampers is specified by four parameters $(r, \gamma, w$ and $k_{off})$, which can be estimated based on published experiments (1–3). The value $r = 1$ µm/s is estimated based on the observed rate of cable growth in vivo. GFP-labelled Smy1 proteins are seen to pause at the bud neck for about a second in wild type cells and so we estimate $k_{off} = 1/s$ for the rate of Smy1 falling off of the formins.

The myosin-aided delivery rate of Smy1 to the formin, leads to a length dependent on rate $k_{on}(l) = wl$. We estimate the value of the parameter $w$ using the observed number of myosin+Smy1 complexes on the cable. If we model the actin cable as a polymer with $l$ subunits, at every subunit we can consider all the processes by which the myosin+Smy1 complexes arrive and depart the particular subunit. In steady state, the number of complexes arriving and departing need to balance. In particular, myosin+Smy1 can either reach the $x^{th}$ subunit ($1 < x < l$) diffusively from the cell cytosol with a rate $k_{on}^0$ (which is proportional to the concentration of Smy1 proteins), or by translocating from the $x - 1$ subunit, with a rate $v$. We assume that the motors do not fall off the polymer and therefore the only way that they leave the $x^{th}$ subunit is by translocating to subunit $x + 1$. At steady state, the number of complexes arriving and departing the $x^{th}$ subunit are equal and therefore the steady state number is $N(x) = \frac{x \, k_{on}^0}{v}$. Using this quantity we can compute the total number of motors (myosin+Smy1 complexes) on the polymer (or cable) by summing over all subunits i.e. $N_{tot} = \sum_{x=0}^{l} N(x) = \frac{k_{on}^0}{v} \frac{l(l+1)}{2}$ (2).

The rate of delivery of Smy1 to the formin at the barbed end is equal to the number of complexes that translocate from the $l^{th}$ subunit to the formin, i.e. $k_{on}(l) = vN(l) = lk_{on}^0$; therefore $k_{on}^0$ is equal to the previously defined parameter $w$. We can solve for $k_{on}^0$, to obtain the relation $N_{tot} = w\frac{L(L+L_0)}{2L_0 V}$, where $V = vL_0$, is the myosin velocity in units of microns per second, and $L = l\, L_0$ is the cable length in microns; $L_0 = 2.7$ nm is the size of an actin subunit in the cable. In the in-vivo experiments, about 5 Smy1+myosin complexes are observed moving at a rate 3.5 μm/s towards the budneck, hence for the purpose of our calculations, $N_{tot} = 5$, $L = 5$ μm, and $V = 3.5$ μm/s which yields $w = 0.004$ s$^{-1}$.

We use these three parameters and the expression for mean cable length in Equation 2 to obtain a value of the fourth parameter, the disassembly rate $\gamma$. By equating the mean cable length to 5 microns (i.e. the diameter of the yeast cell), and using the parameter values listed above, we estimate $\gamma = 0.12$ μm/s or 45 subunits/s.

*4. Active transport of monomers to the site of assembly*

*4.1 Solving the master equations*

The master equation governing the growth of a filament in this mechanism is given by

$$\frac{dP(l,t)}{dt} = \frac{r'}{l-1}P(l-1) + \gamma P(l+1) - \frac{r'}{l}P(l) - \gamma P(l). \quad (32)$$

We will use the scheme of recursion to solve this equation. Note that the master equation blows up at $l = 0$, hence we impose $P(0) = 0$, to avoid this issue. At steady state, substituting $l = 1$ in Equation 32, we obtain, $P(2) = \frac{r'}{\gamma} P(1)$. Similarly, for $l = 2$, we get

$P(3) = \frac{1}{2!} \left(\frac{r'}{\gamma}\right)^2 P(1)$, and for $l = 3$, we get $P(4) = \frac{1}{3} \left(\frac{r'}{\gamma}\right) P(3) = \frac{1}{3!} \left(\frac{r'}{\gamma}\right)^3 P(1)$. Proceeding in a

similar fashion, we obtain $P(l) = \frac{1}{l!} \left(\frac{r'}{\gamma}\right)^l P(1)$, and we can use normalization i.e. .

$\sum_{l=0}^{\infty} P(l) = 1$ to obtain $P(1) = e^{-\frac{r'}{\gamma}}$.

## 5. Depolymerizers

### 5.1 Flux of the motors at the end of the filament

In Figure 9A in the main text, we define $N(x)$ as the number of motors on the $x^{th}$ subunit. Balancing the flux of motors at the $x^{th}$ cell we obtain,

$$vN(x - 1) + k_{on} = vN(x). \qquad (33)$$

In Equation 33, the flux at the $x^{th}$ subunit is equal to the incoming flux of motors on the $(x - 1)^{th}$ subunit and the motors arriving on the polymer from bulk diffusion. At the first subunit, there is no incoming flux; the only motors on the subunit are there through diffusion. Hence we obtain, $N(1) = \frac{k_{on}}{v}$ as $N(0) = 0$. In Equation 33, for $x = 2$, we find $vN(1) + k_{on} = vN(2)$. Substituting $N(1)$ in Equation 33, we get $N(2) = \frac{2 k_{on}}{v}$. Similarly, for x=3, we obtain, $N(3) = \frac{3 k_{on}}{v}$ and so on. Thus solving Equation 33 recursively, we obtain $N(x) = \frac{x k_{on}}{v}$. In other words, the number of motors at a particular subunit depends on the distance from the negative end of the microtubule.

### 5.2 Solving the master equation

The master equation for depolymerizers is given by

$$\frac{dP(l,t)}{dt} = r\,P(l-1) - rP(l) + \gamma' \times (l+1)P(l+1) - \gamma' \times l\,P(l). \qquad (34)$$

As is shown in Equation 34, in order to have a filament of length $l$, the filament can either grow from $l-1$ by adding a subunit with a rate $r$ or depolymerize from $l+1$ with a length dependent rate $\gamma' \times (l+1)$. Note that these terms add to the probability $P(l,t)$, probability that the filament has length $l$ at time $t$. Alternatively the filament of length $l$ add a subunit and become $l+1$ with a rate $r$ or lose a subunit with a length dependent depolymerization rate $\gamma l$. These terms reduce the probability $P(l,t)$.

From Equation 34 for $l = 0$, we get $\frac{dP(0,t)}{dt} = r\,P(-1) - rP(0) + \gamma' P(1)$. Since $l$ is length of the polymer, it cannot be negative, hence $P(-1) = 0$. Thus $\frac{dP(0,t)}{dt} = \gamma' P(1) - rP(0)$. At steady state, $\frac{dP(0,t)}{dt} = 0$. Thus we obtain $P(1) = \frac{r}{\gamma'} P(0)$.

In Equation 34, for $l = 1$, $rP(0) + 2\gamma' P(2) = (r + \gamma')P(1)$. Substituting the expression for $P(0)$ we obtain an expression for $P(2)$ in terms of $P(0)$, namely $r\,P(0) + 2\gamma' P(2) = (r + \gamma')\frac{r}{\gamma'} P(0)$, which simplifies to $P(2) = \frac{1}{2}\left(\frac{r}{\gamma'}\right)^2 P(0)$. Similarly we can obtain an expression for $P(3)$ by substituting $l = 2$ in the Equation 34, i.e. $P(3) = \frac{1}{3}\frac{1}{2}\left(\frac{r}{\gamma'}\right)^3 P(0)$.

Thus Equation 34 can be recursively solved to obtain $P(l) = \left(\frac{r}{\gamma'}\right)^l \frac{1}{l!} P(0)$.

*5.3 Equation for the mean*

We multiply the master equation in Equation 29 by $l$ and sum over all possible lengths, namely

$$\sum_{l=0}^{\infty} \frac{d}{dt} l\, P(l,t) = \sum_{l=0}^{\infty} r\, l\, P(l-1) - \sum_{l=0}^{\infty} r\, l\, P(l) + \sum_{l=0}^{\infty} \gamma'\, l\, (l+1) P(l+1) - \sum_{l=0}^{\infty} \gamma' l^2\, P(l). \quad (35)$$

Let's first consider the first term $\sum_{l=0}^{\infty} r\, l\, P(l-1)$. As defined before, $P(-1) = 0$. Then the first term can be written as $\sum_{l=1}^{\infty} r\, l\, P(l-1)$. Now making a change of variable where $m = l - 1$, we get

$$\sum_{l=1}^{\infty} r\, l\, P(l-1) = \sum_{m=0}^{\infty} r(m+1) P(m). \quad (36)$$

Now let's consider the third term $\sum_{l=0}^{\infty} \gamma'\, l\, (l+1) P(l+1)$. As defined before, $P(-1) = 0$. Then this term can be written as $\sum_{l=-1}^{\infty} \gamma'\, l\, (l+1)\, P(l+1)$. Now making a change of variable where $m = l + 1$, we get

$$\sum_{l=-1}^{\infty} \gamma'\, l\, (l+1)\, P(l+1) = \sum_{m=0}^{\infty} \gamma'(m-1) m\, P(m). \quad (37)$$

Also, $\sum_{l=0}^{\infty} \frac{d}{dt} l\, P(l,t) = \frac{d}{dt} \sum_{l=0}^{\infty} l\, P(l,t)$. Substituting Equation 35 and 36 in Equation 34, we get

$$\frac{d}{dt} \sum_{l=0}^{\infty} l\, P(l,t) = \sum_{m=0}^{\infty} r(m+1) P(m) - \sum_{l=0}^{\infty} r\, l\, P(l) + \sum_{m=0}^{\infty} \gamma'(m-1) m\, P(m)$$

$$- \sum_{l=0}^{\infty} \gamma' l^2\, P(l). \quad (38)$$

Equation 37 simplifies to

$$\frac{d}{dt}\sum_{l=0}^{\infty} l\, P(l,t) = r\sum_{m=0}^{\infty} mP(m) + r\sum_{m=0}^{\infty} P(m) - r\sum_{l=0}^{\infty} l\, P(l) + \gamma\sum_{m=0}^{\infty} m^2 P(m)$$

$$-\gamma'\sum_{m=0}^{\infty} m\, P(m) - \sum_{l=0}^{\infty} \gamma' l^2\, P(l). \quad (39)$$

Using the definition of moments, i.e. $\sum_{l=0}^{\infty} l^n P(l,t) = \langle l^n \rangle$, and the normalization condition $\sum_{l=0}^{\infty} P(l,t) = 1$, we get $\frac{d}{dt}\langle l\rangle = r\langle l\rangle + r - r\langle l\rangle + \gamma'\langle l^2\rangle - \gamma'\langle l\rangle - \gamma'\langle l^2\rangle$, which simplifies to $\frac{d}{dt}\langle l\rangle = r - \gamma'\langle l\rangle$. Note that this is the equation for time dependent average length, i.e. $\langle l\rangle = \langle l(t)\rangle$

Integrating the equation of $\frac{d}{dt}\langle l\rangle$ from $l_0$ to $l$ we get, $\langle l(t)\rangle = \frac{r}{\gamma} + e^{-\gamma' t}\left(l_0 - \frac{r}{\gamma'}\right)$, where $l_0$ is the starting filament length at $t=0$. At steady state ($t \to \infty$), the average length is $\left(\frac{r}{\gamma'}\right)$ (Figure S1).

### 5.4 Effect of finite number of binding sites

In our simple calculation, we assumed that the concentration of motors is small enough, so the capture rate of motors on the microtubule is just $k_{on}^0(s)$ where $k_{on}^0(s) = k_{on}'\,[motors]$. But in reality this rate will depend on the number of binding sites on the microtubule, given by $k_{on}(s) = k_{on}^0\left(1 - \frac{N(x)}{N_{max}}\right)$, where $N(x)$ is the linear density of motors on a microtubule lattice and $N_{max}$ is the maximum number of motors that a microtubule can accommodate. The change in the number of motors will be then given by

$$\frac{dN(x)}{dt} = v\, N(x-1) + k_{on} - v\, N(x). \quad (40)$$

We want to find out how does having a limited number of binding sites changes $N(x)$. As we will see, if the number of available binding sites are large, then this will lead us to the expression we obtained before, namely $N(x) = \frac{k_{on}^0 x}{v}$. Taylor expanding $N(x-1)$ leads to

$$N(x-1) = N(x) - \frac{dN}{dx} + \frac{1}{2}\left(\frac{dN}{dx}\right)^2 + \cdots. \quad (41)$$

Truncating $N(x-1)$ to first order, we can substitute in Equation 35 to obtain

$$\frac{dN}{dt} = -v\frac{dN}{dx} + k_{on}. \quad (42)$$

At steady state, Equation 41 becomes $v\frac{dN}{dx} = k_{on}$. Now we substitute our new expression for $k_{on}(s)$ to get

$$\frac{dN(x)}{dx} = \frac{k_{on}^0}{v}\left(1 - \frac{N(x)}{N_{max}}\right). \quad (43)$$

Now integrating both sides, we get $\int_0^{N(x)}\left(\frac{dN(x)}{\left(1-\frac{N(x)}{N_{max}}\right)}\right) = \frac{k_{on}^0}{v}\int_0^x dx$, which yields,

$\log\left(\frac{N_{max}-N(x)}{N_{max}}\right) = -\frac{k_{on}^0 x}{v\, N_{max}}$. Rearranging the terms, we get $\frac{N_{max}-N(x)}{N_{max}} = e^{-\frac{k_{on}^0 x}{v\, N_{max}}}$, which finally becomes

$$N(x) = N_{max}\left(1 - e^{-\frac{k_{on}^0 x}{v\, N_{max}}}\right). \quad (44)$$

Equation 44 reduces to the expression we got before, i.e. $N(x) = N_{max} \frac{k_{on}^0 x}{v N_{max}} = \frac{k_{on}^0 x}{v}$ when $\frac{k_{on}^0}{v N_{max}} \ll 1$. Hence, as long as $k_{on}^0 \ll v N_{max}$ our simple assumption of $k_{on}^0(s) = k_{on}'[motors]$ is reasonable.

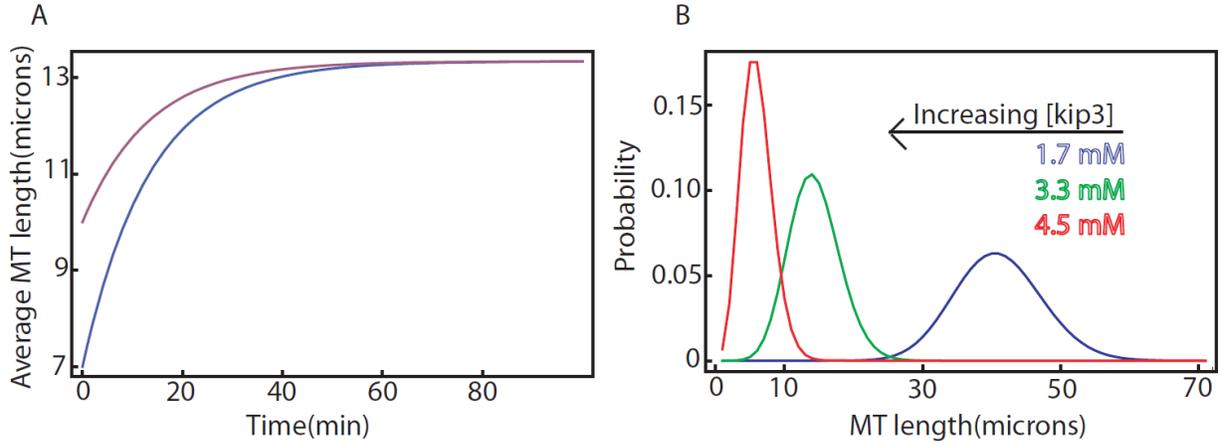

**Figure S1: Predictions for the antenna model of depolymerizers** (A) Time-dependent average length of a microtubule for $l_0 = 7$ μm (in blue) and $l_0 = 10$ μm (in purple), for $r = 1$ μm/min and $\gamma = 0.075$/min. (B) Prediction for the steady state length distribution for different concentrations of Kip3 motors i.e. [Kip3]= 1.7 mM(blue), [Kip3]= 3.3 mM(green) and [Kip3]= 4.5 mM(red). An increase in [Kip3] corresponds to an increase in disassembly rate of the microtubule, and thus we observe that the distributions shift towards lower microtubule length.

## 6. Severing

### 6.1 Solving the master equation

The master equation describing the evolution of $P(l, t)$ in time, for the mechanism of severing is given by

$$\frac{dP(l,t)}{dt} = rP(l-1) - rP(l) + s \sum_{i=l+1}^{\infty} P(i) - s \times (l-1)P(l). \quad (45)$$

This equation is different from the ones described so far. As shown in Equation 41, in order to have a filament of length $l$, the filament can either grow from $l - 1$ by adding a subunit with a rate $r$ or shrink from any filament having a length larger than $l$ by getting severed with a rate $s$. These terms add to the probability $P(l, t)$, that the filament has length $l$ at time $t$. Alternatively, the filament of length $l$ can either get severed $l - 1$ ways with a rate $s$ or add a subunit and become $l + 1$ with a length dependent rate $r$. These terms reduce the probability $P(l, t)$. The first and third terms, thus represent the inflow of probability to and second and fourth terms represent the outflow of probability from the state $l$.

At steady state, the probability does not change with time. In other words, $\frac{dP(l,t)}{dt} = 0$. Hence from Equation 45, we get $P(l)(r + (l-1)s) = s\sum_{i=l+1}^{\infty} P(i) + P(l-1)r$. Using the normalization condition for the probability $P(l)$, $\sum_{i=1}^{\infty} P(i) = \sum_{i=1}^{l} P(i) + \sum_{i=l+1}^{\infty} P(i) = 1$, we obtain

$$P(l)(r + (l-1)s) = s\left(1 - \sum_{i=1}^{l} P(i)\right) + P(l-1)r. \quad (46)$$

Adding the term $sP(l)$ on both sides of the Equation 2, we obtain

$$P(l)(r + ls) = s\left(1 - \sum_{i=1}^{l-1} P(i)\right) + P(l-1)r. \quad (47)$$

For $l = 1$, $P(1)(r + s) = s + P(0)r$. Using the convention $P(0) = 0$, in Equation 43 we find, $P(1) = \frac{s}{r+s}$. For $l = 2$, similarly we find

$$P(2) = \frac{1}{r + 2s}s(1 - P(1)) + P(1)r. \quad (48)$$

Substituting $P(1)$ in Equation 49, we obtain $P(2) = \frac{2\,rs}{r+2s}\left(\frac{1}{r+s}\right)$. Dividing the numerator and denominator by $s^2$, we get $P(2) = \left(\frac{2\frac{r}{s}}{\frac{r}{s}+2}\right)\left(\frac{1}{\frac{r}{s}+1}\right)$. Continuing with the same logic, for $l = 3$ we find $P(3) = \frac{1}{r+3s}s\left(1 - (P(1) + P(2))\right) + P(2)r$. Substituting $P(1)$ and $P(2)$ in this equation, we obtain $P(3) = \frac{3\,r^2 s}{(r+3s)(r+2s)(r+s)}$. Dividing the numerator and denominator by $s^3$, we obtain $P(3) = \frac{3\left(\frac{r}{s}\right)^2}{\left(\frac{r}{s}+3\right)\left(\frac{r}{s}+2\right)\left(\frac{r}{s}+1\right)}$. Thus the master equations in Equation 45 can be recursively solved to give a closed form solution of the form

$$P(l) = \frac{l\rho^{l-1}}{(\rho+1)(\rho+2)..(\rho+l)}, \quad (49)$$

where $\rho = \frac{r}{s}$.

*6.2 Estimation of mean length compared to exact result*

In the case of severing, the disassembly rate (i.e., rate of subunit loss) is length dependent, and can be approximated as $\gamma(l) \sim sl \times \frac{l}{2} = \frac{sl^2}{2}$ (See main text Section 2.3.2). In other words, severing leads to a quadratic dependence of the disassembly rate on the length of the filament (Figure 10C). It is possible to get a rough estimate of the average steady state filament length by equating the disassembly rate $\gamma(l)$ with the assembly rate $r$, i.e. $\langle l \rangle = \sqrt{\frac{2\,r}{s}}$ (Figure S2 in purple). The exact solution of the mean filament length is given by the Equation 29 in Section 2.3.2.1 (Figure S2 in blue).

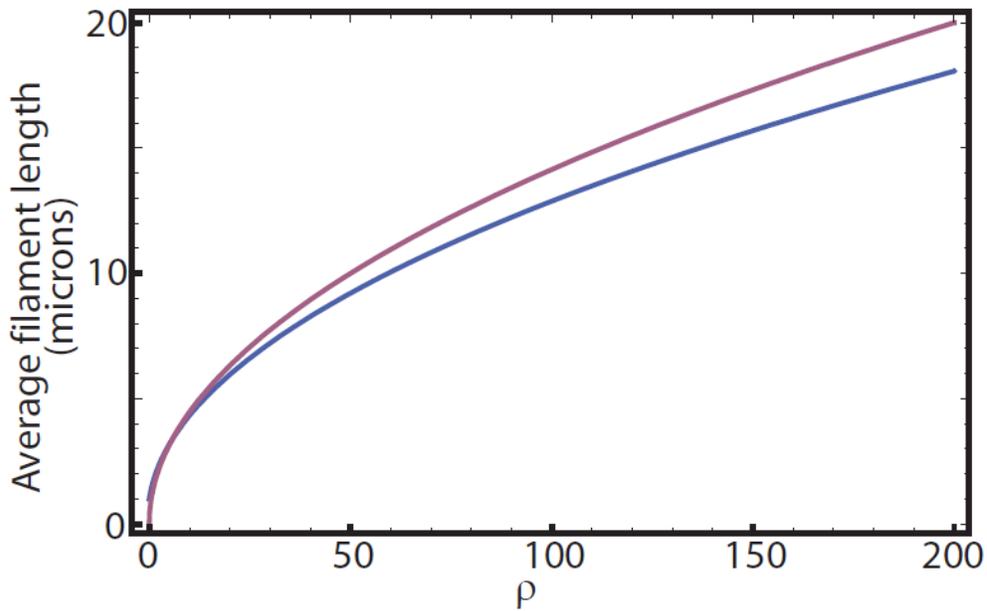

**Figure S2: Average filament length from severing.** This plot compares the average filament length obtained exactly by solving for the steady state distribution (in blue) and by assuming a disassembly rate having a quadratic dependence on the length of the filament (in purple) as a function the severing parameter $\rho = r/s$. The rough estimate using the expression $\langle l \rangle = \sqrt{2r/s}$ is larger than the exact mean evaluated by Equation 43 after $\rho \sim 30$.